\documentclass[12pt,a4paper]{article}
\usepackage[utf8]{inputenc}
\usepackage[DIV13]{typearea}
\usepackage{ragged2e}
\usepackage{calligra,amsmath,amsfonts,mathrsfs,amssymb,bbm,textcomp,bbding}
\usepackage{slashed,cancel,units}
\usepackage[usenames,dvipsnames]{xcolor}
\usepackage{cite}
\usepackage{hyperref}
\hypersetup{colorlinks=true,urlcolor=Magenta,anchorcolor=blue,citecolor=blue,filecolor=blue,
            linkcolor=Magenta,menucolor=blue, linktocpage=true,pdfproducer=medialab}
\usepackage[english]{babel}
\usepackage{catchfile}
\usepackage{indentfirst}
\usepackage{multirow}
\usepackage{pifont}
\usepackage{graphics,graphicx}
\usepackage{enumerate}
\usepackage{booktabs}
\usepackage{bm}
\usepackage{hyperref}
\usepackage{tikz-feynman}
\usepackage[yyyymmdd,hhmmss]{datetime}
\usepackage{amssymb}
\usepackage{amsfonts}
\usepackage{amsmath}
\usepackage{bbm}
\usepackage[scriptsize]{subfigure}
\usepackage{graphicx}
\usepackage{indentfirst}
\usepackage{tikz-feynman}
\usepackage[all]{nowidow}
\usepackage{slashed}
\usepackage{textcomp}
\usepackage{epstopdf}
\usepackage{float}
\usepackage{caption}
\usepackage{enumerate}
\usepackage{mathrsfs}
\usepackage{mathtools}
\usepackage{hhline}
\usepackage{array}
\usepackage{color}
\usepackage{relsize}
\usepackage{bbding}
\usepackage{pifont}
\usepackage{wasysym}
\usepackage{amssymb}
\usepackage{palatino}
\usepackage{pdfpages}
\usepackage[normalem]{ulem}
\newcommand{\email}[1]{\href{mailto:#1}{\tt #1}}

\numberwithin{equation}{section}

\usepackage{catchfile}
\newcommand{\getenv}[2][]{%
  \CatchFileEdef{\temp}{"|kpsewhich --var-value #2"}{}%
  \if\relax\detokenize{#1}\relax\temp\else\let#1\temp\fi}
\getenv[\USER]{USER}

\newcommand{\LL}{\mathscr{L}}

\newcommand{\cO}{\mathcal{O}}

\newcommand{\derp}{\partial}
\newcommand{\hc}{\text{h.c.}}
\newcommand{\ov}[1]{\overline{#1}}
\newcommand{\nn}{\nonumber}
\newcommand{\mueV}{\ \mu\text{eV}}
\newcommand{\meV}{\ \text{meV}}
\newcommand{\eV}{\ \text{eV}}
\newcommand{\keV}{\ \text{keV}}
\newcommand{\TeV}{\ \text{TeV}}
\newcommand{\GeV}{\ \text{GeV}}

\newcommand{\be}{\begin{equation}}
\newcommand{\ee}{\end{equation}}
\definecolor{vierde}{rgb}{0.0, 0.5, 0.0}

\newcommand{\blue}[1]{{\color{blue} #1 \color{black}}}

\newcommand{\nef}{N_\text{eff}}
\newcommand{\dnef}{\Delta N_\text{eff}}

\begin{document}
\renewcommand*{\thefootnote}{\fnsymbol{footnote}}
\begin{titlepage}

\vspace*{-1cm}
\flushleft{FTUAM-20-5}
\hfill{IFT-UAM/CSIC-20-34}
\\[1cm]

\begin{center}
\blue{\bf\LARGE 
Production of Thermal Axions \\ \vspace{0.3cm} across the ElectroWeak Phase Transition
 }
 \\[4mm]
\centering

\end{center}
\vskip 0.5  cm
\begin{center}
Fernando~Arias-Arag\'on$^{a,b)}$ 
\footnote{E-mail: \email{fernando.arias@uam.es}},
Francesco~D'Eramo$^{c,d)}$
\footnote{E-mail: \email{francesco.deramo@pd.infn.it} },\\[3mm]
Ricardo~Z. Ferreira$^{e,f)}$ 
\footnote{E-mail: \email{ricardo.zambujal@su.se}},
Luca~Merlo$^{a,b)}$ 
\footnote{E-mail: \email{luca.merlo@uam.es}}, and 
Alessio~Notari$^{g)}$ 
\footnote{E-mail: \email{notari@fqa.ub.edu}}
\vskip .7cm
{\footnotesize
\centerline{$^{a)}$ \it Instituto de F\'isica Te\'orica UAM/CSIC, Calle Nicol\'as Cabrera 13-15, Cantoblanco E-28049 Madrid, Spain}\vspace*{0.2cm}
\centerline{$^{b)}$ \it Departamento  de  F\'{\i}sica Te\'{o}rica,  Universidad  Aut\'{o}noma  de  Madrid, Cantoblanco  E-28049  Madrid,  Spain}\vspace*{0.2cm}
\centerline{$^{c)}$ \it Dipartimento di Fisica ed Astronomia, Universit\`a di Padova, Via Marzolo 8, 35131 Padova, Italy}\vspace*{0.2cm}
\centerline{$^{d)}$ \it INFN, Sezione di Padova, Via Marzolo 8, 35131 Padova, Italy}\vspace*{0.2cm}
\centerline{$^{e)}$ \it Nordita, KTH Royal Institute of Technology and Stockholm University,} 
\centerline{\it Roslagstullsbacken 23, SE-106 91 Stockholm, Sweden}\vspace*{0.2cm}
\centerline{$^{f)}$ \it Institut de F\'isica d'Altes Energies (IFAE) and The Barcelona Institute of Science and Technology (BIST),} 
\centerline{\it Campus UAB, 08193 Bellaterra, Barcelona} \vspace*{0.2cm}
\centerline{$^{g)}$ \it Departament de F\'isica Qu\`antica i Astrofis\'ica \& Institut de Ci\`encies del Cosmos (ICCUB),}
\centerline{\it Universitat de Barcelona, Mart\'i i Franqu\`es 1, 08028 Barcelona, Spain}
}
\end{center}

\vskip 0.5cm

\begin{abstract}
\justify Light axions can potentially leave a cosmic background, just like neutrinos. We complete the study of thermal axion production across the electroweak scale by providing a smooth and continuous treatment through the two phases. Focusing on both flavor conserving and violating couplings to third generation quarks, we compute the amount of axions produced via scatterings and decays of thermal bath particles. We perform a model independent analysis in terms of axion effective couplings, and we also make predictions for specific microscopic QCD axion scenarios. This observable effect, parameterized as it is conventional by an effective number of additional neutrinos, is above the $1\sigma$ sensitivity of future CMB-S4 surveys. Moreover, if one assumes no large hierarchies among dimensionless axion couplings to standard model particles, future axion helioscopes will provide a complementary probe for the parameter region we study. 

\end{abstract} 

\end{titlepage}
\setcounter{footnote}{0}

\pdfbookmark[1]{Table of Contents}{tableofcontents}
\tableofcontents

\renewcommand*{\thefootnote}{\arabic{footnote}}


\section{Introduction}
\label{sec:intro}

Understanding the absence of CP violation by strong interactions, an issue known as the strong CP problem, is a serious challenge for the Standard Model (SM) of particle physics. This remarkable invariance is unexpected since there are two potential sources of CP violation in the QCD Lagrangian,
\be
\LL_{\rm QCD} \supset \dfrac{\alpha_s}{8\pi} \, \theta \, G^a_{\mu\nu} \, \widetilde{G}^{a \mu\nu} - 
\left[\sum_{ij} \overline{q_{L i}} \, M_{ij} \, q_{R j} +{\rm h.c.} \right] \ .
\label{eq:LQCDCP}
\ee
Here, $\alpha_s = g_s^2 / (4\pi)$ is the QCD fine structure constant, $q_{L,R i}$ are the quarks fields with $i$ a flavor index, $G^a_{\mu\nu}$ and $\widetilde{G}^{a \mu\nu} = \epsilon^{\mu\nu\rho\sigma} G^a_{\rho\sigma} / 2$ are the QCD field strength tensor and its dual, respectively. An overall phase in the quark mass matrix $M_{ij}$, ${\rm arg}({\rm det}(M)) \neq 0$, provides an additional source of CP violation in the quark sector beyond the CKM phase. One can transfer the full amount of CP violation on the first operator with $\theta$ replaced by $\bar{\theta} \equiv \theta - {\rm arg}({\rm det}(M))$, and the non observation of a neutron electric dipole moment puts the spectacular bound $\bar{\theta} \leq 1.3 \times10^{-10}$~\cite{Baker:2006ts,Afach:2015sja}.

An elegant solution was proposed in the late 70's by Peccei and Quinn (PQ)~\cite{Peccei:1977hh,Peccei:1977ur}. They introduced a new Abelian symmetry $U(1)_\text{PQ}$, dubbed PQ symmetry, with two key features: anomalous with respect to the $SU(3)_c$ color gauge group, and spontaneously broken at a scale $f_a$. The low-energy residual is a pseudo-Nambu-Goldstone boson (PNGB) $a$, known as the axion~\cite{Wilczek:1977pj,Weinberg:1977ma}, which acquires the anomalous coupling to gluons
\be
\LL_{\rm axion} \supset \dfrac{\alpha_s}{8\pi} \, \frac{a}{f_a} \,  G^a_{\mu\nu} \, \widetilde{G}^{a \mu\nu}  \ .
\label{eq:LQCDCP}
\ee
This equation, valid before the axion mixes with the $\eta$ and $\pi^0$ mesons, defines $f_a$. Non-perturbative QCD effects generate an axion potential, and a theorem due to Vafa and Witten~\cite{Vafa:1984xg} ensures a CP conserving minimum. Moreover, the axion potential leads to the general relation for its mass~\cite{Bardeen:1978nq,diCortona:2015ldu}
\be
m_a=5.70(6)(4)\mueV\left(\dfrac{10^{12}\GeV}{f_a}\right) \ . 
\label{massequation}
\ee
The first error is due to the uncertainty in the up-down quark mass ratio whereas the second one is due to uncertainties in low energy couplings. Axion couplings are proportional to $1 / f_a$, and $f_a$ is bound by astrophysical and terrestrial searches~\cite{Anastassopoulos:2017ftl,Jaeckel:2015jla,Brivio:2017ije,Bauer:2017ris,Irastorza:2018dyq}, spanning the range $f_a \gtrsim10^6 - 10^9\GeV$: the axion must be light and weakly-coupled (scenario dubbed as the invisible axion~\cite{Kim:1979if,Shifman:1979if,Zhitnitsky:1980tq,Dine:1981rt}). The energy density stored in the axion field can account for dark matter (DM) for values of $f_a$ allowed experimentally~\cite{Marsh:2015xka}. 

The focus of this work is on a different and distinct cosmological imprint: scattering and/or decay of particles in the primordial plasma produce relativistic axions~\cite{Turner:1986tb,Masso:2002np}. Current bounds on $f_a$ implies that $m_a$ must be roughly below the eV scale. Axions produced at early times are still relativistic at matter-radiation equality and, for $m_a\ll \cO(0.1)\eV$ as we consider by neglecting the axion mass, also around recombination. In this case, they would manifest themselves as an additional contribution to the amount of radiation at the time of CMB formation. Upcoming CMB-S4 surveys~\cite{Abazajian:2016yjj,Abazajian:2019eic} will improve bounds on this quantity, historically parameterized as an effective number of neutrino species $\nef$, and can potentially discover a deviation from the SM. The forecasted sensitivity allows to detect the effects of a relativistic species which decoupled at high temperatures, as high as the ElectroWeak Phase Transition (EWPT), making this a new probe of high-energy physics. Motivated by forthcoming data, recent works revisited axion production through various channels and the resulting prediction for $\nef$~\cite{Brust:2013xpv,Salvio:2013iaa,Baumann:2016wac,Ferreira:2018vjj,DEramo:2018vss}. Furthermore, this could be a complementary probe of the axion interpretation for the excess in the number of electron recoil events observed recently by the XENON1T experiment~\cite{Aprile:2020tmw} as highlighted by Ref.~\cite{Arias-Aragon:2020qtn}.

There are, broadly speaking, two classes of axion interactions with visible matter
\be
\LL_{\rm axion-int} \supset  \frac{1}{f_a} \left[  a\, c_X \dfrac{\alpha_X}{8\pi} \, X^{a \mu\nu}\widetilde{X}^a_{\mu\nu} + 
\partial_\mu a \; c_\psi \overline{\psi} \gamma^\mu \psi \right]    \ .
\label{eq:axionINTgeneric}
\ee
Operators with gauge bosons $X=\{G,\,W, \,B\}$, present if the PQ symmetry is anomalous under the associated gauge group, are suppressed by a loop factor. We need a coupling to gluons in order to solve the strong CP problem, and we set $c_G = 1$ consistently with Eq.~\eqref{eq:LQCDCP}. Anomalies under the electroweak group are possible but not mandatory. We consider the high-energy theory where SM fermions, $\psi = \left\{Q_L, u_R, d_R, L_L, e_R \right\}$, have well defined gauge quantum numbers  and their interactions with the axion preserve the shift symmetry $a \, \rightarrow \, a + {\rm const}$. Other dimension 5 interactions can be redefined away as explained later in the text.

Axion production is efficient when the interaction rate exceeds the Hubble rate $H$. The latter, assuming an early universe dominated by radiation with temperature $T$, scales as $H \propto T^2 / M_{\rm Pl}$. Hot axions can be produced either via scatterings or decays. Interactions with gauge bosons, the first kind in Eq.~\eqref{eq:axionINTgeneric}, cannot mediate decays. Coupling to SM fermions, the second kind in Eq.~\eqref{eq:axionINTgeneric}, could in principle be responsible for production via decays if the fermion bilinear couples fields belonging to different generations. In other words, we need flavor violation in order to have production via unsuppressed tree-level bath particles decays to axions. 

Regardless of the production details, the highest value for $\nef$ is reached when axions achieve thermalization with the thermal bath. The resulting abundance in this case depends only on the plasma temperature when they lose thermal contact, and its value is suppressed by the total number of the entropic degrees of freedom $g_{*s}$ at decoupling. If this happens above the EWPT then the resulting $\nef$ is barely within the reach of future surveys~\cite{Brust:2013xpv}. 

Rates for scattering mediated by interactions with SM gauge bosons scale at high temperatures as $\Gamma_X \simeq \alpha_X^3  T^3 / f_a^2$, and these processes are never in thermal equilibrium at the EWPT for $f_a\gtrsim \cO(10^8)\GeV$~\cite{Masso:2002np,Salvio:2013iaa}. If axions never thermalize, the prediction for $\nef$ is sensitive to the initial abundance that is presumably set at the stage of reheating after inflation; an accurate calculation requires a treatment of thermal effects at high temperature~\cite{Salvio:2013iaa}. Either way, the associated $\nef$ is at the edge of what we can test. 

Once considering interactions with fermions, for flavor conserving couplings, the only way to produce hot axions is via scattering. We consider $2 \rightarrow 2$ collisions, and these processes always involve two SM fermions and one SM boson besides the axion itself.

At temperatures above the EWPT, we have two options for the SM boson involved. On the one hand, it can be any of the four real component of the Higgs doublet $H$~\footnote{Namely the Higgs boson and what would become the longitudinal components of the weak gauge bosons below the EWPT scale.} and the scattering rate in this case scales as $\Gamma_{\psi/H} \simeq c_\psi^2 y^2_\psi T^3 / f_a^2$~\cite{Salvio:2013iaa} with $y_\psi$ the SM fermion Yukawa coupling. For heavy SM fermions, this is larger by a factor $c_\psi^2 y_\psi^2/\alpha^3_X$ compared to scattering mediated by the axion-gauge boson vertex. On the other hand, it can be a transverse gauge boson. The scattering rate in this case is proportional to the mass of the fermion, since there is a chirality flip needed in the process, and this contribution is vanishing because all fermions are massless above the EWPT.

The SM boson involved in the scattering with fermions can be a gauge field only below the EWPT. The associated rate scales as  $\Gamma_{\psi/X} \simeq \alpha_X c_\psi^2 m_\psi^2 T / f_a^2$ for temperatures above the fermion mass, where  $m^2_\psi$ reflects the fermion chirality flip mentioned in the paragraph above, and it is exponentially suppressed at lower temperatures. In this case, at temperatures above the fermion mass, the scattering rate grows with the temperature slower than the Hubble rate and thus axion production is saturated when the ratio between  interaction and expansion rates is maximal. This happens at temperatures around the fermion mass, and such a ratio is approximately $\Gamma_{\psi/X}/H|_{T=m_\psi}\approx  \alpha_X  c_\psi m_\psi M_{Pl} / f_a^2$. If this quantity is larger than ${\cal O}(1)$ thermalization is achieved, and the final abundance is not affected by our ignorance about the thermal history (assuming reheating above the weak scale) and possible new degrees of freedom and/or interactions at high energy.

Decays of SM fermions provide an additional axion production channel, often the dominant one, if we have flavor violating couplings. The interaction rate is given by the rest frame width of the decaying fermion times a Lorentz dilation factor accounting for the bath kinetic energy, and it scales as $\Gamma_\psi \simeq c_\psi^2 m_\psi^4 / (f_a^2 T)$. Axion production is saturated at temperatures around the fermion mass also in this case, and we achieve thermalization if the condition $\Gamma_{\psi}/H|_{T=m_\psi}\approx c_\psi^2 m_\psi M_{Pl} / f_a^2 $ is satisfied.

After this comparison among different production channels, we decide to focus on axion production mediated by its interactions with SM fermions. We analyze processes with third generation quarks. Production via leptons has been studied in Ref.~\cite{DEramo:2018vss}, and such an axion abundance can alleviate the current tension in the measurement of the Hubble parameter~\cite{Bernal:2016gxb}. A full calculation via the first two quark generations would require a careful treatment of the QCD phase transition (QCDPT) and it is beyond the scope of this work. Previous studies have considered production well above~\cite{Salvio:2013iaa} and well below~\cite{Ferreira:2018vjj} the EWPT. We improve earlier treatments by providing a continuous and smooth prediction for $N_\text{eff}$ across the EWPT.

We introduce the theoretical framework in Sec.~\ref{sec:Lagrangian}, and we describe axion effective interactions considering both flavor conserving and violating couplings. We collect in Sec.~\ref{sec:AxionProduction} all the processes contributing to axion production, and we provide explicit expressions for cross sections and decay widths. In particular, we compute cross section both above and below the EWPT and we match them at this threshold. We feed Boltzmann equations with these quantities and we solve them numerically, presenting predictions for $\nef$ as a function of the fermion couplings in Sec.~\ref{sec:results}. We consider both effective interactions as well as explicit UV constructions leading to flavor conserving  couplings. Remarkably, our predictions are within the reach of future CMB surveys inside the low-$f_a$ part of the experimentally allowed region. It is to be noted that, as we discuss in our conclusions in Sec.~\ref{sec:conclusions}, in the absence of no big hierarchy in the dimensionless coefficient describing the coupling to photons, we find that the same parameter space will be probed by future terrestrial searches.


\section{Axion Effective Interactions}
\label{sec:Lagrangian}

Axion interactions with SM fields can be written compactly as follows
\be
\LL^{(a)} = \LL^{(a)}_\text{gauge} + \LL^{(a)}_\text{matter} \ ,
\ee
where $\LL^{(a)}_\text{gauge}$ and $\LL^{(a)}_\text{matter}$ describe couplings with SM gauge bosons and matter fields, respectively. The entire focus of our work is on axion couplings with SM quarks. However, there are usually relations among different axion interactions once one considers UV complete models. For this reason, we provide an overview of all axion couplings and we summarize their bounds in App.~\ref{app:bounds} in order to visualize which parameter space region is not excluded experimentally. 

The axion has anomalous couplings to gauge bosons 
\be
\LL^{(a)}_\text{gauge}=- 
\dfrac{a}{f_a}
\left(\dfrac{\alpha_s}{8\pi}G_{\mu\nu}^a\widetilde{G}^{a\,\mu\nu}+c_W\dfrac{\alpha_W}{8\pi}W_{\mu\nu}^a\widetilde{W}^{a\,\mu\nu}+c_B\dfrac{\alpha_Y}{8\pi}B_{\mu\nu}\widetilde{B}^{\mu\nu}\right)\,,
\ee
where $\widetilde{W}$ and $\widetilde{B}$ are defined as $\widetilde{G}$ below Eq.~\eqref{eq:LQCDCP}. These operators should be interpreted as the effects of the presence of any fermion that couples to the axion and are associated to quantum level contributions. As already mentioned in the Introduction, the gluon term does not present any free coefficient, in contrast with the EW terms, in order to match with the traditional definition of $f_a$. Once we integrate-out weak scale states and heavy quarks, the  Lagrangian contains axion couplings to only gluons and photons
\be
\LL^{(a)}_\text{gauge}\supset-
\dfrac{a}{f_a}
\left(\dfrac{\alpha_s}{8\pi}G_{\mu\nu}^a\widetilde{G}^{a\,\mu\nu}+c_{a\gamma\gamma}\dfrac{\alpha_\text{em}}{8\pi}F_{\mu\nu}\widetilde{F}^{\mu\nu}\right)\,,
\label{EffectiveGaugeBosonLag}
\ee
where $c_{a\gamma\gamma}= c_B \, \cos\theta_W^2 + c_W \, \sin\theta_W^2$, being $\theta_W$ the Weinberg angle. This expression is valid above the scale where strong interactions confine and therefore before the axion mixes with the $\eta$ and $\pi^0$ mesons. However, experimental searches probe the axion-photon coupling at much lower energy scales and therefore this mixing is to be taken into account. We define this coupling as follows:
\be
g_{a\gamma\gamma}\equiv\dfrac{\alpha_\text{em}}{2\pi}\dfrac{1}{f_a}\Big(c_{a\gamma\gamma}-1.92(4)\Big)\,,
\label{caggcagammagammadef}
\ee
where the second term in the parenthesis is the model-independent contribution arising from the above mentioned axion mixing with the $\eta'$ and $\pi^0$ mesons ~\cite{Georgi:1986df,Kaplan:1985dv,Srednicki:1985xd,Bardeen:1986yb,diCortona:2015ldu}. 

We present matter couplings for the case of quarks, the discussion is analogous if we consider leptons.  Axion couplings to quarks can be expressed in different field basis and physical results cannot depend on such a choice. However, the statement that the axion couples without flavor violation is not true in an arbitrary basis. We specify axion couplings to quarks in the ``primed basis'' defined in App.~\ref{app:basis} where the fields appearing in the Lagrangian are the $SU(2)_L$ quark doublets $Q'_L$, and the $SU(2)_L$ singlets $u'_R$ and $d'_R$, and where Yukawa interactions take the form of Eq.~\eqref{YukawasChoicePrime}. 

We distinguish between two cases, and we begin from flavor conserving axion-quark interactions
\be
\LL^{(a)}_\text{matter} \supset  \LL^{(a)}_{\partial-\text{F.C.}} =
\dfrac{\partial_\mu a}{f_a} \sum_{i=1}^3 \left(c_Q\,\ov{Q'_{Li}}\gamma^\mu Q'_{Li} + c_u\,\ov{u'_{Ri}}\gamma^\mu u'_{Ri} + c_d\,\ov{d'_{Ri}}\gamma^\mu d'_{Ri}\right)\,,
\label{eq:aqLag}
\ee
where the free coefficients $\left\{c_Q, c_u, c_d \right\}$ are typically of the same order of magnitude. The universality of quark-axion couplings guarantees that no flavor-changing interactions arise when moving to the quark mass basis. We can see it explicitly after performing the rotation to get mass eigenstates given in Eq.~\eqref{eq:massbasisrotation}, the unitarity of the CKM matrix ensures that in the mass eigenbasis the fermion couplings are still flavor conserving.

The most general flavor violating part of the Lagrangian above the EWPT can be written in an analogous way to the flavor conserving one as follows
\be
\LL^{(a)}_\text{matter} \supset \LL^{(a)}_{\partial-\text{F.V.}} = 
\frac{\partial_\mu a}{f_a}\displaystyle\sum_{i,j} \left(c^{(ij)}_Q\,\ov{Q'_{Li}}\gamma^\mu Q'_{Lj} + c^{(ij)}_u \, \ov{u'_{Ri}}\gamma^\mu u'_{Rj} + 
c^{(ij)}_d \, \ov{d'_{Ri}}\gamma^\mu d'_{Rj} \right) \ ,
\label{eq:aqLag2}
\ee
where the matrices of coefficients $\left\{c^{(ij)}_Q, c^{(ij)}_u, c^{(ij)}_d \right\}$ have a generic structure in flavor space. Unless we tune the entries of these matrices consistently with CKM factors, couplings are still flavor off-diagonal once we go to the mass eigenstate basis.

We complete this overview on axion couplings by discussing the remaining options. The case of coupling to leptons is analogous, and Ref.~\cite{DEramo:2018vss} exploited their cosmological consequences. Besides interactions with leptons, no other matter couplings can be present in the Lagrangian as an independent operator. The Higgs-axion interaction
\be
i\dfrac{\derp_\mu a}{f_a}H^\dag \overset\leftrightarrow{D^\mu} H \ ,
\ee
where $H^\dag \overset\leftrightarrow{D^\mu} H\equiv H^\dag (D^\mu H) - (D^\mu H)^\dag H$ is redundant at lowest order in $1/f_a$ as can be shown via a field redefinition. Moreover, axion couplings to pseudo-scalar fermion currents such as
\be
i \dfrac{a}{f_a}\ov{Q'_L}\,H\,d'_R\,,
\ee
can be proved to be also redundant.


\section{Axion Production Processes}
\label{sec:AxionProduction}

Multiple processes contribute to the production of hot axions in the early universe, and we list all of them in this section. Binary scatterings control production for the flavor conserving case since decays are loop and CKM suppressed. We provide the associated scattering cross sections above and below the weak scale, and we match our results across the EWPT. If axion couplings are flavor violating then tree-level decays dominate the production rate. We give here the associated decay widths. 

Axion couplings to quarks are a crucial ingredient for our calculations, and we remark how we define them in the ``primed basis'' where the SM Yukawa interactions take the form in Eq.~\eqref{YukawasChoicePrime}. One of our main goals is to provide a smooth treatment of production through the EWPT, hence it is convenient to work in the mass eigenbasis. 

\subsection{Cross sections}

We start from flavor conserving axion couplings defined in Eq.~\eqref{eq:aqLag} and quantified by the scale $f_a$ and the three dimensionless coefficients $\left\{c_Q, c_u, c_d \right\}$. As it turns out, scattering cross sections depend only on two linear combinations of them. This can be checked through explicit calculations or via a change of basis. We perform the following rotations where we redefine quark fields by an axion-dependent phase
\be
Q'_{Li} \, \rightarrow e^{i c_Q \frac{a}{f_a}} Q'_{Li}  \ , \qquad\qquad  
u'_{Rj} \, \rightarrow e^{i c_u \frac{a}{f_a}} u'_{Rj} \ , \qquad  \qquad
d'_{Rj} \, \rightarrow e^{i c_d \frac{a}{f_a}} d'_{Rj} \ .
\label{eq:chiralrot}
\ee
These chiral rotations modify several couplings in the Lagrangian. First, they are anomalous and the dimensionless coefficients of axion couplings to gauge bosons in Eq.~\eqref{EffectiveGaugeBosonLag} are affected; as already stated above we do not consider these interactions for our processes and we do not need to worry about this effect. Second, we generate new axion derivative couplings equal and opposite to the ones in Eq.~\eqref{eq:aqLag} once we plug the new quark fields defined above in the kinetic terms. Thus axion derivative couplings are not present anymore in the Lagrangian. Third, and crucially for us, the axion field appears in the Yukawa interactions after we plug these field redefinitions into the SM Yukawa Lagrangian, whose explicit expression is given in Eq.~\eqref{YukawasChoicePrime}, and we find
\be
- \LL^{(a)}_{Y - \text{F.C.}} = e^{i (c_u - c_Q) \frac{a}{f_a}} \ov{Q'_L}\,\widetilde{H}\,\widehat{Y}^u\,u'_R + e^{i (c_d - c_Q) \frac{a}{f_a}} \ov{Q'_L}\,H\,V_\text{CKM}\,\widehat{Y}^d\,d'_R+\hc \ .
\label{eq:aqLagYukawaBasis}
\ee
As anticipated, although there are three different couplings in the theory only two linear combinations of them can appear in scattering amplitudes
\begin{align}
\label{eq:ReducedAxionCouplings1} c_t \equiv \, - c_Q + c_u \ , \\
\label{eq:ReducedAxionCouplings2} c_b \equiv \, - c_Q + c_d \ .
\end{align}
We label them with the top and bottom quark because we only focus on the third quark generation as explained in the Introduction. The hatted matrices $\hat{Y}^{u,d}$ are diagonal in flavor space, and axion interactions are flavor conserving once we switch to the mass eigenbasis via the rotations given in Eq.~\eqref{eq:massbasisrotation}.

\subsubsection*{Scattering cross sections above EWPT}

\begin{table}  \centering 
\begin{tabular}{|c|c|c|}
\hline
\multicolumn{3}{|c|}{\textbf{Axion Production Above EWSB}} \\ \hline \hline
Process  & CP Conjugate & $\sigma_{ij \rightarrow k a} \times 64 \pi f_a^2$ \\
\hline \hline
&&\\[-4mm]
$t \bar{t} \rightarrow \chi_0 a$ & $t \bar{t} \rightarrow \chi^c_0 a$ & $c_t^2y_t^2$ \\
$b \bar{b} \rightarrow \chi_0 a$ &$b \bar{b} \rightarrow \chi^c_0 a$ & $c_b^2 y_b^2$ \\
$t \bar{b} \rightarrow \chi_+ a$ &$b \bar{t} \rightarrow \chi_- a$ & $c_t^2y_t^2 + c_b^2 y_b^2$ \\[1mm] \hline
&&\\[-4mm]
$t \chi_0 \rightarrow t a$ & $\bar{t} \chi^c_0 \rightarrow \bar{t} a$ & $c_t^2y_t^2$ \\
$t \chi_0^c \rightarrow t a$ & $\bar{t} \chi_0 \rightarrow \bar{t} a$ & $c_t^2y_t^2$ \\
$b \chi_0 \rightarrow b a$ & $\bar{b} \chi^c_0 \rightarrow \bar{b} a$ & $c_b^2y_b^2$ \\
$b \chi_0^c \rightarrow b a$ & $\bar{b} \chi_0 \rightarrow \bar{b} a$ & $c_b^2y_b^2$ \\
$t \chi_- \rightarrow b a$ & $\bar{t} \chi_+ \rightarrow \bar{b} a$ & $c_t^2y_t^2 + c_b^2y_b^2$ \\
$b \chi_+ \rightarrow t a$ & $\bar{b} \chi_- \rightarrow \bar{t} a$ & $c_t^2y_t^2 + c_b^2y_b^2$ \\[1mm]  \hline
\end{tabular}
\caption{\em Scatterings producing axions above the EWPT. In the first two columns we list the process and its CP conjugate. They have the same cross section, listed on the third column.} 
\label{tab:above}
\end{table}

We focus on third generation quarks $\left\{t_L, b_L, t_R, b_R\right\} = \left\{u_{L 3}, d_{L 3}, u_{R 3},  d_{R 3}\right\}$ where we assign new names to left- and right-handed fields. In order to write explicitly their interactions, we parameterize the complex components of the Higgs doublet as follows
\be
H = \begin{pmatrix}
 \chi_+\\[2mm]
 \chi_0
\end{pmatrix} \ , \qquad \qquad \qquad  
\widetilde{H}\equiv i \sigma_2 (H^\dag)^T =  
\begin{pmatrix}
 \chi_0^c \\[2mm]
 - \chi_{-}
\end{pmatrix} \ ,
\label{eq:HiggsField}
\ee
where we define $\chi_{-} \equiv \chi_+^\dag$ and $\chi_0^c \equiv \chi_0^\dag$. Once we focus on third generation quarks and we consider the Lagrangian in Eq.~\eqref{eq:aqLagYukawaBasis} in the mass eigenbasis, namely without the CKM matrix, we find the following axion interactions
\be
\begin{split}
- \LL^{(a)}_{Y - \text{F.C.}} = &\phantom{+}  y_t \, e^{i c_t \frac{a}{f_a}} \, \left[ \chi_0^c \, \ov{t_L} t_R - \chi_- \,  \ov{b_L} t_R \right] + 
y_b \, e^{i c_b \frac{a}{f_a}} \, \left[ \chi_+ \, \ov{t_L} b_R + \chi_0 \,  \ov{b_L} b_R \right] + \\ 
&
+y_t \, e^{- i c_t \frac{a}{f_a}} \, \left[ \chi_0 \, \ov{t_R} t_L - \chi_+ \,  \ov{t_R} b_L \right] + 
y_b \, e^{- i c_b \frac{a}{f_a}} \, \left[ \chi_- \, \ov{b_R} t_L + \chi_0^c \,  \ov{b_R} b_L \right] \ .
\label{eq:axioncouplingsYukawa}
\end{split}
\ee
The processes we are interested in have only one axion field in the external legs, thus we can Taylor expand the exponential functions appearing in the above Lagrangian and only keep terms up to the first order.

In the unbroken electroweak phase, the Higgs vev is vanishing and all particles are massless. We want to consider processes producing one axion particle in the final state thus the most general binary collisions involve two fermions fields. The other boson in the process can be either a component of the Higgs doublet or a SM gauge boson. However, if we look at the axion interactions in Eq.~\eqref{eq:axioncouplingsYukawa} we see that only the former is possible. There is no $2 \rightarrow 2$ scattering with SM gauge bosons; this is manifest in the basis we choose to describe axion couplings. Alternatively, if we insisted on working in the basis where axion is derivatively coupled to SM fermions the amplitude for a $2 \rightarrow 2$ is vanishing as it requires a fermion chirality flip that is not possible in the absence of a mass term for the fermion itself. 

We only have processes with the components of the Higgs doublet in Eq.~\eqref{eq:HiggsField}. The two fermions in the scattering can be either both in the initial state or one in the initial state and the other one in the final state. We classify all possible cases according to where fermions appear. If we consider the first case, we have fermion/antifermion annihilations producing and axion and any of the components of the complex Higgs doublet. The possible processes are listed in the first block of Tab.~\ref{tab:above}; we show the associate CP conjugate on the same row, and we correctly account for both in our numerical analysis. Another possibility is to have just one fermion in the initial state, and the other particle would be a component of the Higgs doublet. The associated processes are listed in the second block of Tab.~\ref{tab:above}. For each process we also provide the scattering cross section. Our contribution proportional to $y_t^2$ agrees with what was found in Ref.~\cite{Salvio:2013iaa}.

\subsubsection*{Scattering cross sections below EWPT}

\begin{table} \centering 
\begin{tabular}{|c|c|c||c|c|c|}
\hline
\multicolumn{6}{|c|}{\textbf{Axion Production Below EWSB}} \\ \hline \hline
Process  & CP Conjugate & $\sigma_{ij \rightarrow k a}$ &Process  & CP Conjugate & $\sigma_{ij \rightarrow k a}$ \\
\hline \hline
&&&&&\\[-4mm]
$t \bar{t} \rightarrow g a$ & Same  & \multirow{2}{*}{Eq.~\eqref{eq:qqga}} & $t g \rightarrow t a$ & $\bar{t} g \rightarrow \bar{t} a$ &\multirow{2}{*}{Eq.~\eqref{eq:qgqa}} \\
$b \bar{b} \rightarrow g a$ & Same  && $b g \rightarrow b a$ & $\bar{b} g \rightarrow \bar{b} a$ &\\
$t \bar{t} \rightarrow h a$ & Same  & \multirow{2}{*}{Eq.~\eqref{eq:qqha}} & $t h \rightarrow t a$ & $\bar{t} h \rightarrow \bar{t} a$ & \multirow{2}{*}{Eq.~\eqref{eq:qhqa}} \\
$b \bar{b} \rightarrow h a$ & Same  & &$b h \rightarrow b a$ & $\bar{b} h \rightarrow \bar{b} a$ & \\
$t \bar{t} \rightarrow Z a$ & Same  & Eq.~\eqref{eq:ttZa}& $t Z \rightarrow t a$ & $\bar{t} Z \rightarrow \bar{t} a$ & Eq.~\eqref{eq:tZta} \\ 
$b \bar{b} \rightarrow Z a$ & Same  & Eq.~\eqref{eq:bbZa}& $b Z \rightarrow b a$ & $\bar{b} Z \rightarrow \bar{b} a$ & Eq.~\eqref{eq:bZba}\\ 
\multirow{2}{*}{$t \bar{b} \rightarrow W^+ a$} & \multirow{2}{*}{$b \bar{t} \rightarrow W^- a$} & \multirow{2}{*}{Eq.~\eqref{eq:tbWa}}&$t W^- \rightarrow b a$ & $\bar{t} W^+ \rightarrow \bar{b} a$  &Eq.~\eqref{eq:tWba}\\
&&&$b W^+ \rightarrow t a$ & $\bar{b} W^- \rightarrow \bar{t} a$ & Eq.~\eqref{eq:bWta} \\  \hline
\end{tabular}
\caption{\em Scatterings producing axions below the EWPT. We give the process (left column), the CP conjugate (central column) and the reference to the equation with the explicit analytical expression for the scattering cross section.} 
\label{tab:below}
\end{table}

Once the electroweak symmetry is broken, the Higgs field gets a vacuum expectation value (vev) which gives mass to SM particles. We work in unitarity gauge where the Higgs field is parameterized by the following field coordinates
\be
H = \begin{pmatrix}
0\\[2mm]
 \frac{v + h}{\sqrt{2}}
\end{pmatrix} \ , \qquad \qquad \qquad  
\widetilde{H}\equiv i \sigma_2 (H^\dag)^T =  
\begin{pmatrix}
  \frac{v + h}{\sqrt{2}} \\[2mm]
 0
\end{pmatrix} \ ,
\ee
where $v$ and $h$ are the vev and the physical Higgs boson, respectively. In such a gauge, the three remaining (Goldstone) components of the Higgs doublet are eaten up by the massive Z and W bosons. The mass spectrum as a function of the Higgs vev results in
\be
\left\{ m_W, m_Z, m_h, m_f \right\} = \left\{ \frac{g}{2}, \frac{\sqrt{g^2+{g'}^2}}{2}, \sqrt{\frac{\lambda}{2}}, \frac{y_f}{\sqrt{2}}  \right\} v \ .
\ee
Here, $g$ and $g'$ are the $SU(2)_L$ and $U(1)_Y$ gauge couplings, respectively. The Higgs quartic coupling $\lambda$ is normalized in such a way that the potential term is $\lambda (H^\dag H)^2$.

We list in  Tab.~\ref{tab:below} all processes contributing to axion production in the phase where the electroweak symmetry is broken. As done already above, we provide also the CP conjugate process as well as the scattering cross section. The explicit expressions are too lengthy to be displayed directly in the table and we give their explicit analytical expressions in App.~\ref{app:XSbelow}.

\subsubsection*{Matching at the EWPT}

We complete our discussion of production via scattering by showing how processes involving the four components of the Higgs doublet, three of which are the longitudinal components of the Z and W bosons below the EWPT, match at the point of electroweak symmetry breaking. For this reason, the $Z$ and $W$ components involved in the following processes are the longitudinal ones and will be denoted with an index $L$ in the rest of this subsection. In the following, we will take the cross sections for the different processes, express all masses in terms of the Higgs vev and run towards $v\rightarrow 0$. The process that will match are shown in Tab.~\ref{tab:matching}.

\begin{itemize}
\item {\it Neutral annihilations: $t \bar{t} \rightarrow \chi_0 a$, $\quad$ $t \bar{t} \rightarrow \chi^c_0 a$ ; $\quad$ $t \bar{t} \rightarrow h a$, $\quad$ $t \bar{t} \rightarrow Z_L a$.}

When considering the limit in which the Higgs vev vanishes, the cross sections below EWPT coincide exactly with those above, as expected:
\begin{equation}
\sigma_{t \bar{t} \rightarrow \chi_0 a}+\sigma_{t \bar{t} \rightarrow \chi^c_0 a}=\sigma_{t \bar{t} \rightarrow h a}+\sigma_{t \bar{t} \rightarrow Z_L a} = \frac{c_t^2y_t^2}{32\pi f_a^2}.
\end{equation}

\item {\it Neutral scatterings: $t \chi_0 \rightarrow t a$, $\quad$ $t \chi_0^c \rightarrow t a$ ; $\quad$ $t h \rightarrow t a$, $\quad$ $t Z_L \rightarrow t a$. \\ 
$\left.\right.$ \hspace{3.22cm} $\bar{t} \chi_0 \rightarrow \bar{t} a$, $\quad$ $\bar{t} \chi^c_0 \rightarrow \bar{t} a$ ; $\quad$ $\bar{t} h \rightarrow \bar{t} a$, $\quad$ $\bar{t} Z_L\rightarrow \bar{t} a$.}

In this case, the matching can be expressed as:

\begin{equation}
\begin{split}
& \sigma_{t \chi_0 \rightarrow t a} +\sigma_{t \chi_0^c \rightarrow t a}
= \sigma_{t h \rightarrow t a}+\sigma_{t Z_L \rightarrow t a} = \\
= & \, \sigma_{\bar{t} \chi^c_0 \rightarrow \bar{t} a}+\sigma_{\bar{t} \chi_0 \rightarrow \bar{t} a}
=\sigma_{\bar{t} h \rightarrow \bar{t} a}+\sigma_{\bar{t} Z_L\rightarrow \bar{t} a}= \frac{c_t^2y_t^2}{32\pi f_a^2},
\end{split}
\end{equation}

where the CP conjugates give indeed the same contribution.

\item {\it Charged annihilations: $t \bar{b} \rightarrow \chi_+ a$, $\quad$ $b \bar{t} \rightarrow \chi_- a$ ; $\quad$ $t \bar{b} \rightarrow W^+_L a$, $\quad$ $b \bar{t} \rightarrow W^-_L a$.}

Analogously to the neutral case, the charged annihilations can be matched as:

\begin{equation}
\sigma_{t \bar{b} \rightarrow \chi_+ a}+\sigma_{b \bar{t} \rightarrow \chi_- a}=\sigma_{t \bar{b} \rightarrow W^+_L a}+\sigma_{b \bar{t} \rightarrow W^-_L a} = \frac{c_t^2y_t^2+c_b^2y_b^2}{32\pi f_a^2}.
\end{equation}

\item {\it Charged scatterings: $t \chi_- \rightarrow b a$, $\quad$ $b \chi_+ \rightarrow t a$ ; $\quad$ $t W^-_L \rightarrow b a$, $\quad$ $b W^+_L \rightarrow t a$ . \\
$\left.\right.$ \hspace{2.81cm} $\quad$ $\bar{t} \chi_+ \rightarrow \bar{b} a$, $\quad$ $\bar{b} \chi_- \rightarrow \bar{t} a$ ; $\quad$ $\bar{t} W^+_L \rightarrow \bar{b} a$, $\quad$ $\bar{b} W^-_L\rightarrow \bar{t} a$.}

Finally, these set of processes match their cross section as follows:
\begin{equation}
\begin{split}
& \sigma_{t \chi_- \rightarrow b a} + \sigma_{b \chi_+ \rightarrow t a}
=\sigma_{t W^-_L \rightarrow b a}+\sigma_{b W^+_L \rightarrow t a} = \\ 
= & \, \sigma_{\bar{t} \chi_+ \rightarrow \bar{b} a}+\sigma_{\bar{b} \chi_- \rightarrow \bar{t} a}
=\sigma_{\bar{t} W^+_L \rightarrow \bar{b} a}+\sigma_{\bar{b} W^-_L\rightarrow \bar{t} a}
= \frac{c_t^2y_t^2+c_b^2y_b^2}{32\pi f_a^2}.
\end{split}
\end{equation}
\end{itemize}

\begin{table} \centering
\begin{tabular}{|c|c|}
\hline
Processes Above EWPT & Processes Below EWPT\\
\hline \hline
&\\[-4mm]
$t \bar{t} \rightarrow \chi_0 a\, + \,t \bar{t} \rightarrow \chi^c_0 a$ & $t \bar{t} \rightarrow h a\, + \,t \bar{t} \rightarrow Z_L a$ \\
$t \bar{b} \rightarrow \chi_+ a$ & $t \bar{b} \rightarrow W_L^+ a$\\
$b \bar{t} \rightarrow \chi_- a$ & $b \bar{t} \rightarrow W_L^- a$\\
$t \chi_0 \rightarrow t a\, + \,t \chi_0^c \rightarrow t a$ & $t h \rightarrow t a\, + \,t Z_L \rightarrow t a$\\
$\bar{t} \chi_0 \rightarrow \bar{t} a\, + \,\bar{t} \chi^c_0 \rightarrow \bar{t} a$ & $\bar{t} h \rightarrow \bar{t} a\, + \,\bar{t} Z_L \rightarrow \bar{t} a$\\
$t \chi_- \rightarrow b a$ & $t W_L^- \rightarrow b a$\\
$\bar{t} \chi_+ \rightarrow \bar{b} a$ & $\bar{t} W_L^+ \rightarrow \bar{b} a$\\
$b \chi_+ \rightarrow t a$ & $b W_L^+ \rightarrow t a$\\
$\bar{b} \chi_- \rightarrow \bar{t} a$ & $\bar{b} W_L^- \rightarrow \bar{t} a$\\ \hline
\end{tabular}
\caption{\em Scatterings producing axions involving the Higgs doublet above and below the EWPT. We consider the four components of the Higgs doublet above, and we work in unitary gauge below with the Higgs boson $h$ and the longitudinal components $Z_L$ and $W_L$ of the weak bosons.} 
\label{tab:matching}
\end{table}

\subsection{Decay widths}

The crucial ingredient for axion production via quark decays is the Lagrangian with flavor violating interactions whose explicit expression is given in Eq.~\eqref{eq:aqLag2}. This channel is active only below the EWPT because it is kinematically forbidden at higher temperatures where all quarks are massless. For this reason, we find it convenient to rewrite it in terms of Dirac quark fields $u = (u_L \; u_R)$ and $d = (d_L \; d_R)$ in the mass eigenstate basis
\be
\LL^{(a)}_{\partial-\text{F.V.}} = 
\frac{\partial_\mu a}{f_a}\displaystyle\sum_{i,j} \left[ \bar{u}_{i}\gamma^\mu\left( c^{(ij)}_{V_u} + c^{(ij)}_{A_u} \gamma_5\right) u_{j} + 
\bar{d}_{i}\gamma^\mu\left(c^{(ij)}_{V_d} + c^{(ij)}_{A_d} \gamma_5\right)d_{j},\right] \ ,
\label{eq:LagFlavViol}
\ee
where we identify the following combinations
\be
\left\{ c^{(ij)}_{V_u}, c^{(ij)}_{A_u}, c^{(ij)}_{V_d}, c^{(ij)}_{A_d} \right\} = \frac{1}{2} \times
\left\{ c^{(ij)}_u + c^{(ij)}_Q, c^{(ij)}_u - c^{(ij)}_Q, c^{(ij)}_d + c^{(ij)}_Q, c^{(ij)}_d - c^{(ij)}_Q \right\} \ .
\ee
The decay process $q_i\rightarrow q_j a$ and its CP conjugate $\bar{q}_i \rightarrow \bar{q}_ja$, which can happen for both up- and down-type quarks, have the following decay width
\be
\Gamma_{q_i\rightarrow q_ja}=\frac{m_i^3}{16\pi f_a^2} \left({c^{(ij)}_{V_q}}^2+{c^{(ij)}_{A_q}}^2\right) \left(1-\frac{m_j^2}{m_i^2}\right)^3  \ .
\label{eq:DecayRate}
\ee
Here, the ratio $\frac{m_j}{m_i}$ can safely be neglected as it leads to non observable changes in $\dnef$. The decays relevant to our analysis and their CP conjugates are displayed in Tab.~\ref{tab:decays} with their corresponding decay widths.

\begin{table} \centering 
\begin{tabular}{|c|c|c|}
\hline
\multicolumn{3}{|c|}{\textbf{Axion Production Above EWSB}} \\ \hline \hline
Process  & CP Conjugate & $\Gamma_{i\rightarrow ja} \times 16 \pi f_a^2 / m_i^{3}$ \\
\hline \hline
&&\\[-4mm]
$t \rightarrow c a$ & $\bar{t} \rightarrow \bar{c} a$ & ${c^{(tc)}_{V_u}}^2+{c^{(tc)}_{A_u}}^2$ \\
$t \rightarrow u a$ & $\bar{t} \rightarrow \bar{u} a$ & ${c^{(tu)}_{V_u}}^2+{c^{(tu)}_{A_u}}^2$ \\[1mm] \hline
&&\\[-4mm]
$b \rightarrow s a$ & $\bar{b} \rightarrow \bar{s} a$ & ${c^{(bs)}_{V_d}}^2+{c^{(bs)}_{A_d}}^2$\\
$b \rightarrow d a$ & $\bar{b} \rightarrow \bar{d} a$ & ${c^{(bd)}_{V_d}}^2+{c^{(bd)}_{A_d}}^2$ \\[1mm] \hline
\end{tabular}
\caption{\em Quark decays producing axions. In the first two columns we list the process and its CP conjugate, and they both have the same decay widths listed on the third column.} 
\label{tab:decays}
\end{table}


\section{Observable Consequences and Results}
\label{sec:results}

The physical observable of interest in our work is the effective number of neutrinos $\nef$ induced by hot axions. Big bang nucleosynthesis~\cite{Fields:2019pfx} and CMB experiments~\cite{Abazajian:2016yjj,Aghanim:2018eyx} probe this quantity, and we focus on the latter case as it is the most sensitive. Here, we first review briefly how to compute $\nef$ from a given axion production source and then we quantify the $\nef$ generated from all processes analyzed in the previous section.

The effective number of neutrinos is related to the radiation energy density $\rho_\text{rad}$ as
\begin{eqnarray}
\rho_\text{rad} =  \rho_\gamma \left[ 1+ \frac{7}{8} \left(\frac{T_\nu}{T_\gamma}\right)^{4} \nef \right]\, ,
\end{eqnarray}
where $\rho_\gamma$ is the photon energy density. Any relativistic particle with a non-negligible energy density, like neutrinos or axions, will contribute to $\nef$. 
In particular, we are interested in deviations from the $\Lambda$CDM value
\begin{eqnarray}
\Delta \nef \equiv \nef - \nef^{\Lambda \text{CDM}} = \frac{8}{7} \left(\frac{11}{4}\right)^{4/3} \frac{\rho_a}{\rho_\gamma } \ ,
\end{eqnarray}
with $\nef^{\Lambda \text{CDM}}= 3.046$ and $\rho_a$ is the axion energy density.

In order to connect with the numerical solutions of Boltzmann equations, we find it convenient to rewrite $\Delta \nef$ in terms of the comoving axion abundance $Y_a \equiv n_a/s$. Here, $n_a$ is the axion number density and $s= 2\pi^2 g_{*s} T^3/45$ is the entropy density with $g_{*s}$ the number of entropic degrees of freedom. The photon energy density can also be expressed as follows
\be
\rho_\gamma = 2\times \frac{\pi^2}{30} \left(\frac{45\,s}{2\pi^2 g_{*s}}\right)^{4/3}  \ ,
\ee
whereas the axion energy density is related to the number density via 
\be
\rho_a = \frac{\pi^2}{30} \left(\frac{\pi^2 n_a}{\zeta(3)}\right)^{4/3} \, .
\ee
We combine these two equations and find
\begin{eqnarray}
\Delta \nef \simeq 74.85 \, Y_a ^{4/3} \, ,
\end{eqnarray}
where we used the value of $g_{*s}$ at recombination $g_{*s}= 43/11$. 

We determine the asymptotic value of the axion number density by solving the associated Boltzmann equation. The differential equation describing how the axion number density evolves with time reads 
\be
\frac{d}{dt} n_a + 3 H n_a = \left[\sum_S \bar{\Gamma}_S +\sum_D \bar{\Gamma}_D \right] \left( n_a^\text{eq}-n_a \right) \ .
\ee
Here, $H$ is the Hubble parameter quantifying the expansion rate of the universe and the superscript ``eq'' for number density denotes expressions valid when particles are in thermal equilibrium. The two terms on the right hand side denote, respectively, the sum over thermally averaged scattering and decay rates involving the axion and their explicit expressions read
\begin{align}
\bar{\Gamma}_S = & \,  \frac{g_ig_j}{32\pi^4 n_{a}^\text{eq}}T\displaystyle\int_{s_{min}}^\infty\textstyle{ds}\ \frac{\lambda\left(s,m_i,m_j\right)}{\sqrt{s}}\ \sigma_{ij\rightarrow ka}\left(s\right)K_1\left(\frac{\sqrt{s}}{T}\right) \ , \\
\label{eq:DecayColOp} \bar{\Gamma}_D = & \, \frac{n_i^\text{eq}}{n_a^\text{eq}} \frac{K_1\left(\frac{m_i}{T}\right)}{K_2\left(\frac{m_i}{T}\right)}\Gamma_{i\rightarrow ja} \ . 
\end{align}
The function $\lambda$ is defined as follows
\be
\lambda\left(x,y,z\right) \equiv \left[x-\left(y+z\right)^2\right]\left[x-\left(y-z\right)^2\right] \ ,
\ee
whereas the minimum center of mass energy is $s_{min}=\text{Max}\left(\left(m_i+m_j\right)^2,m_k^2\right)$. The general expression for the equilibrium number density is
\begin{equation}
n^\text{eq}_i = \frac{m_i^2}{2\pi^2}TK_2\left(\frac{m_i}{T}\right),
\end{equation}
where $g_i$ stands for the degrees of freedom of the particle $i$ and $K_n(z)$ are the modified Bessel function of the second kind.

We find it convenient to switch to dimensionless variables. We define $x = m/T$, where $m$ is taken to be the mass of the heaviest particle in the process, and the Boltzmann equation for $Y_a$ reads
\begin{eqnarray}
sHx \frac{dY_a}{dx} = \left(1- \frac{1}{3} \frac{\ln g_{*s}}{\ln x}\right) \left(\sum_S \gamma_S + \sum_D \gamma_D \right) \left(1- \frac{Y_a}{Y_a^\text{eq}}\right) \ ,
\end{eqnarray}
where $\gamma_{D,S} \equiv n_a^\text{eq} \,\bar{\Gamma}_{D,S}$. We solve now the equation numerically for the different axion production processes. Analytical approximations can be found for the cases below thermal abundance and at large $f_a$, leading to $\dnef\propto f_a^{-8/3}$~\cite{Ferreira:2018vjj}.

\subsection{Model Independent results}

We first perform a model-independent operator analysis. For flavor conserving couplings, we switch on separately the top-axion vertex $c_t$ and the bottom-axion vertex $c_b$, whereas we account for the decay of each quark for flavor violating interactions. 

\begin{figure}
	\centering
	\includegraphics[height=0.32\textheight]{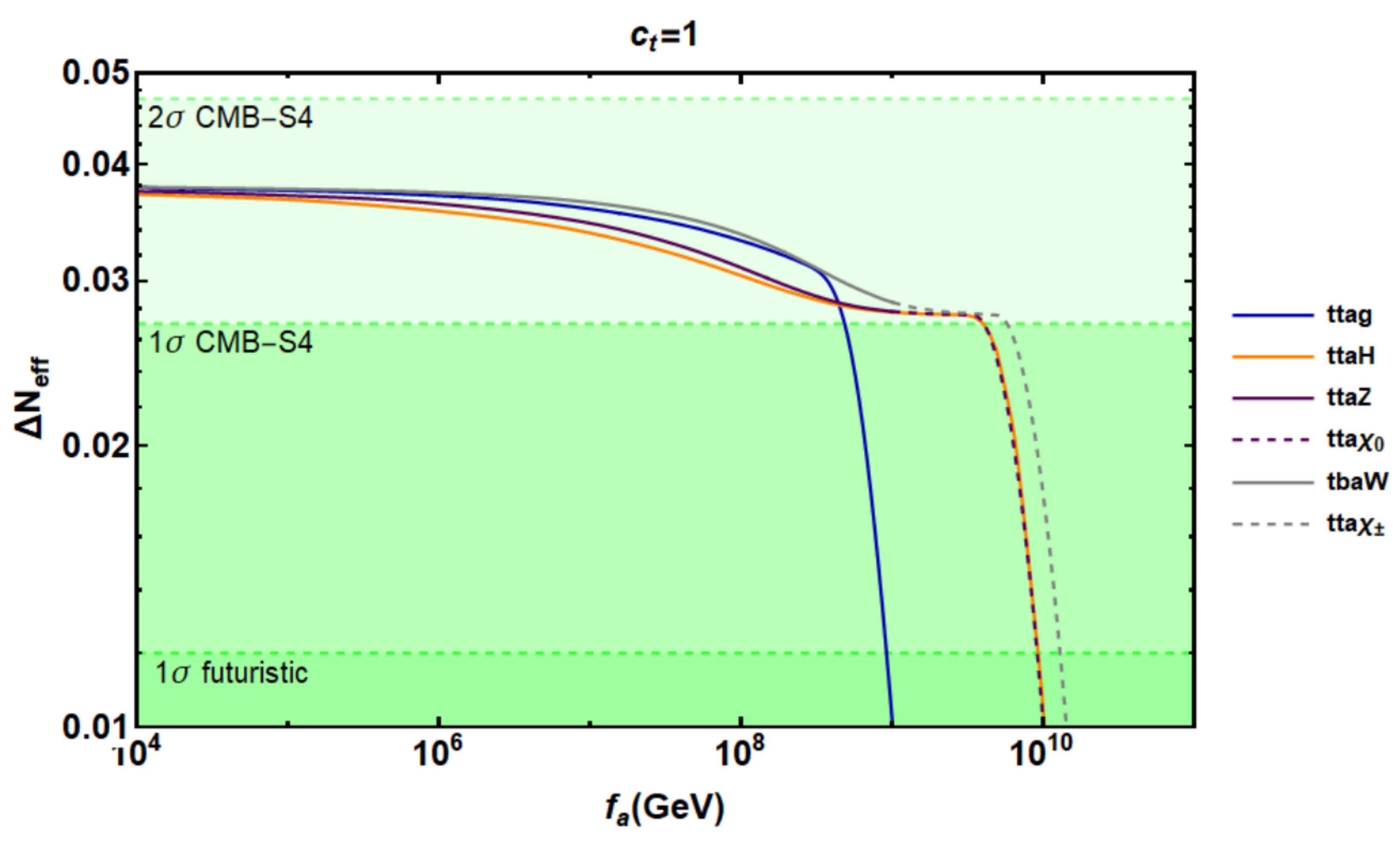}
	\caption{\em Contribution to $\Delta \nef$ from individual binary scatterings with the top quark involved. The dimensionless coupling is set to $c_t = 1$. Each line denotes all processes with the external legs denoted in the legenda. The initial temperature here is set to $T_{I}=10^4$ GeV and the initial axion abundance has been assumed to be zero.}
	\label{fig:FlavConsTop}
\end{figure}

We begin with scatterings and we set $c_i = 1$ throughout this section; this is equivalent to interpreting $f_a$ as $f_a/c_i$ for each specific coupling. One of our main results is a smooth treatment of the EWPT, and this is relevant once one accounts for axion production controlled by processes with the top (anti-)quark on the external legs. We show in Fig.~\ref{fig:FlavConsTop} the contribution to $\Delta \nef$ from each one of these processes as a function of $f_a$. Production via scatterings with gluons is not altered by this threshold. However, processes with longitudinal weak gauge bosons feel this transition since they become massive below the EWPT. Dashed lines correspond to calculations in the electroweak symmetric phase whereas solid lines hold below the EWPT. Our lines for each individual process, with the relevant degrees of freedom at the associated temperature, are indeed smooth across the two phases. 

\begin{figure}
	\centering
	\includegraphics[height=0.32\textheight]{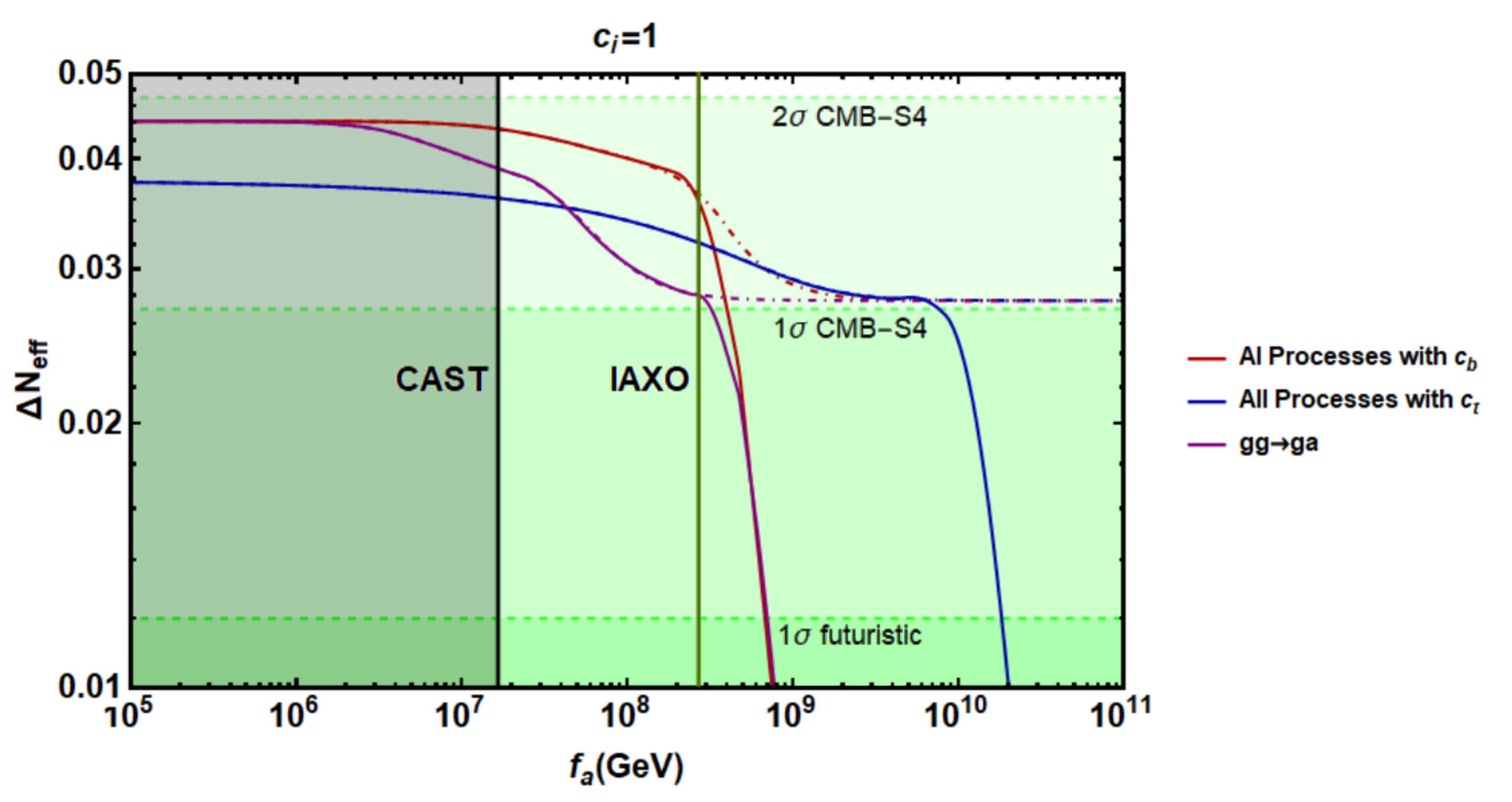}
	\caption{\em Impact on $\Delta \nef$ following an operator-by-operator analysis: for each line, we consider the axion coupling only to one particle $i$, with coupling constant $c_i=1$. The initial temperature here is set to $T_{I}=10^4$ GeV and the initial axion abundance has been assumed to be zero (thermal) for the solid (dot-dashed) lines. The CAST limit and IAXO prospect are shown as a shaded region and a vertical green line respectively, assuming $c_{a\gamma\gamma}=1$.}
	\label{fig:FlavCons}
\end{figure}

The physical observable is actually the combined effects of these individual lines. We add them up and we show our prediction for $\dnef$ in Fig.~\ref{fig:FlavCons} together with the associated quantity from the bottom-axion vertex $c_b$. Solid lines correspond to the extreme case in which the initial temperature $T_I$ ({\it i.e.} the reheating temperature, if the Universe went through a stage of Inflation) was very close to the EWPT, and assuming the initial abundance of axions to be zero at $T=T_I$. The opposite extreme case, dot-dashed lines, correspond to an initial thermal abundance of axions at a given initial temperature above the EWPT. Finally, we show predictions from one particular process which remains the same at all temperatures and whose strength grows with the temperature, the purely gluonic $gg\rightarrow ga$. In order to take it into account, we interpolated the result from Ref.~\cite{Salvio:2013iaa} and assumed that it decreases always with the same power of temperature, extrapolating the results to lower temperatures.

Fig. \ref{fig:FlavCons} shows that the axion can thermalize through the scatterings with the top, below or around the EWPT, for $f_a \lesssim 10^{10}$ GeV even with zero initial axion abundance close to the EWPT. This means that, independently of the initial conditions, it is possible to be above the $1\sigma$ region of CMB-S4. If one assumes an initial thermal abundance, as shown in the same figure with dot-dashed lines, such initial seed automatically gives a signal of about $1\sigma$, as already stressed in previous works \cite{Baumann:2016wac}.

For both choices of initial conditions, the signal increases as $f_a$ lowers, reflecting the fact that the axion decouples at a lower temperature where the number of degrees of freedom in thermal equilibrium $g_*(T_\text{dec})$ is smaller. For $f_a\lesssim 10^9$ GeV the processes involving the axion-bottom coupling become efficient and yields a larger $\Delta \nef$ which roughly saturates at $\Delta \nef(g_*(m_b))$ for $f_a\lesssim 10^8$ GeV. Finally, we note that the axion-gluon scattering channel is always less efficient than the other scattering channels, except for $f_a \lesssim 5\times 10^7$ GeV where it becomes more efficient than the axion-top scatterings.

We also show the constraints from CAST \cite{Arik:2008mq} and the forecasted sensitivity of IAXO \cite{Irastorza:2011gs,Armengaud:2014gea}. Although these experiment probe the axion-photon coupling, we can still compare both forecasts assuming $c_{a\gamma \gamma}=1$. Interestingly, the parameter space probed by IAXO corresponds to the region where $\Delta \nef$ are above the $1\sigma$ level. These multiple detection channels will be very useful in case of a detection.

\begin{figure}
	\centering
	\includegraphics[height=0.32\textheight]{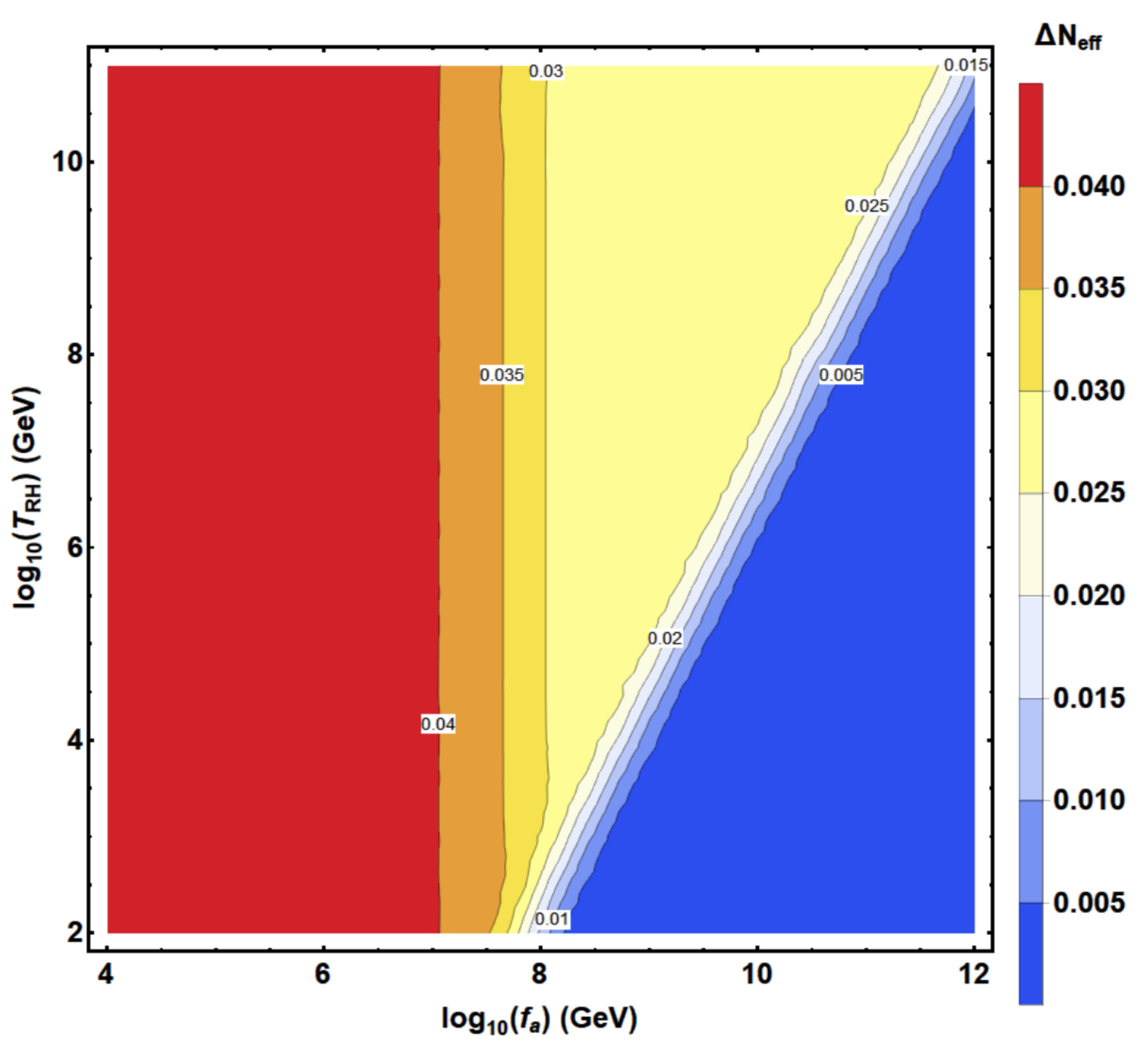}
	\caption{\em $\dnef$ as a function of $f_a$ and the reheating temperature with initial axion abundance set to zero. In this figure only purely gluonic processes have been included.}
	\label{figfavsTRH}
\end{figure}

As discussed in the paragraphs above, the initial conditions for our Boltzmann equation evolution depend on whether axions thermalize or not above the EWPT. For example, for $f_a\sim 10^9$ GeV the axion thermalizes already at $T\sim$ TeV due to the interactions with the Higgs \cite{Salvio:2013iaa}. In general, this depends on the value of the reheating temperature. This interplay between the axion scale $f_a$ and the initial (reheating) temperature, when zero initial abundance is assumed for the axion, can be seen more clearly in Fig.~\ref{figfavsTRH}. Here, we show the dependence of $\Delta \nef$ on the reheating temperature and $f_a$ by considering purely gluonic processes, which must be present in any QCD axion model, following the procedure from before. The figure shows that the axion always thermalizes, independently of $f_a$, as long as the reheating temperature is set high enough.

Turning now to the possibility of having flavor violating couplings, the interactions in Eq.~\eqref{eq:LagFlavViol} lead to the following possible decays below the EWPT
\be
\begin{gathered}
t\to c\,a \ , \qquad\qquad t\to u\,a \ ,\\
b\to s\,a \ , \qquad\qquad b\to d\,a.\\
\end{gathered}
\label{Processes}
\ee
The decays $c\to u\,a$ and $s\to d\,a$ are not taken into consideration, although they should be significant, because the relevant temperatures here should be around the QCD phase transition, where we do not have control on the complicated strongly coupled physics. 

The couplings in Eq.~\eqref{eq:LagFlavViol} also lead to quark annihilation into a $W$ and an axion with non-diagonal flavor transitions: the flavor change may be present in the coupling with the $W$ and/or with the axion. These processes, however, are subdominant with respect to those with only flavor conserving couplings and for these reasons they are not discussed here. The only potentially interesting processes would be $t\,\bar{c}\to Z\,a$ and $t\,\bar{s}\to W\,a$, but the contribution from these processes is of the same order of magnitude as that coming from the processes that involve flavor-conserving couplings, namely $t\,\bar{t}\to Z\,a$ and $t\,\bar{b}\to W\,a$, which were already discussed above. Since the results would be essentially the same, we will not include the analysis of such processes in this work.

\begin{figure*}
	\centering
	\includegraphics[width=0.49\linewidth]{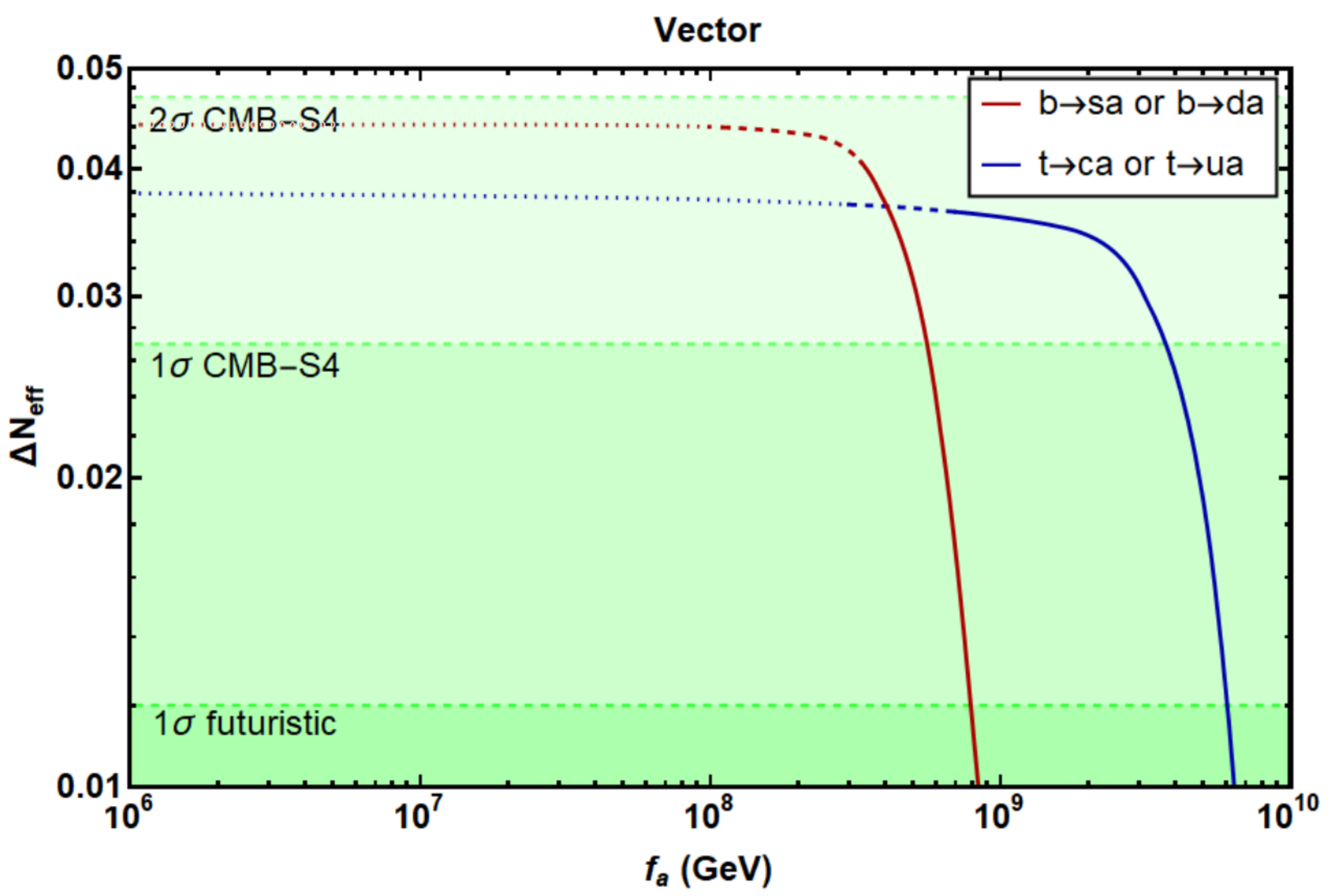}
	\includegraphics[width=0.49\linewidth]{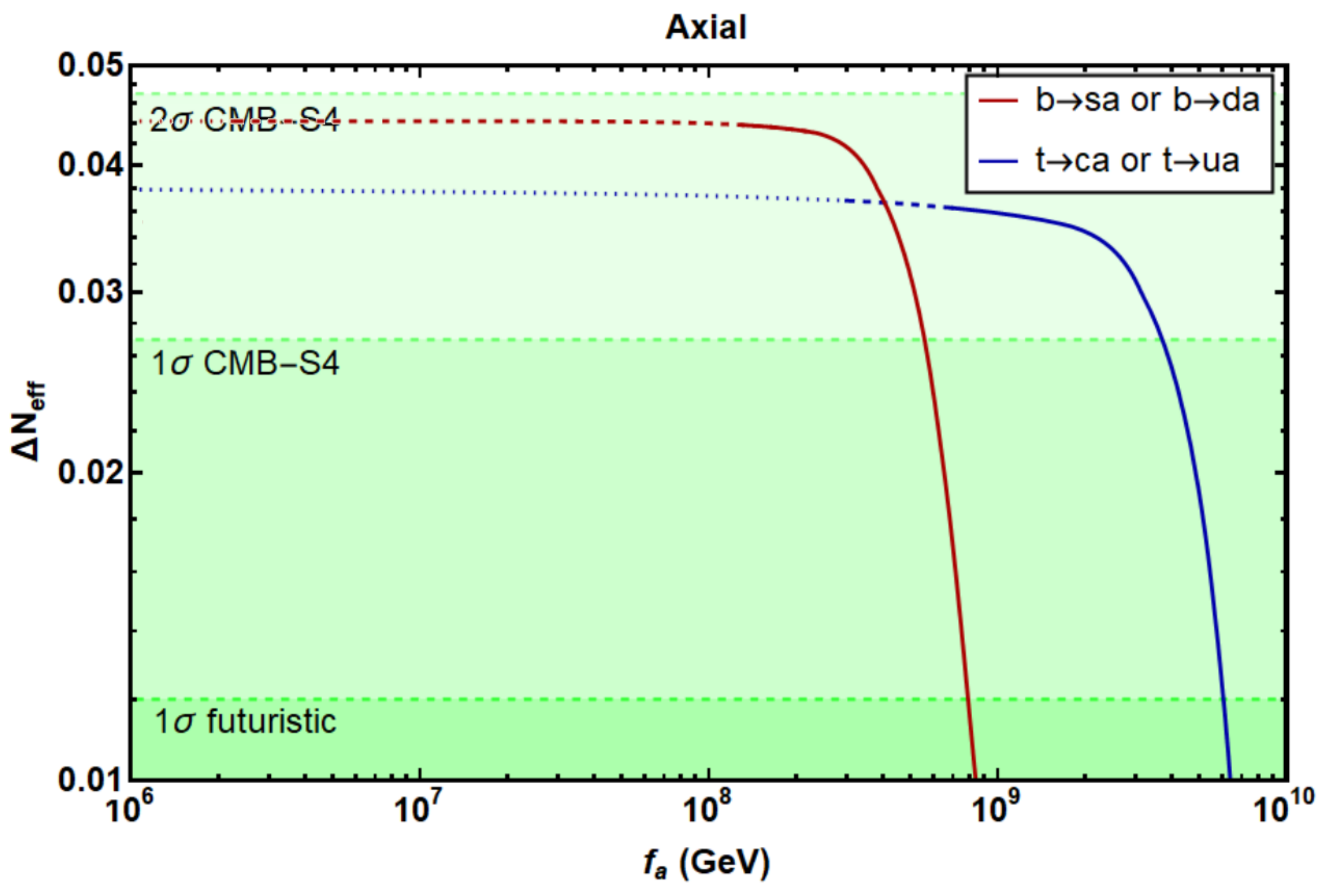}
	\caption{\em Effect of quark decays on $\nef$. The figure on the left (right) assume only vector (axial) couplings, assumed to be equal to one. In these figures solid lines escape all bounds, whereas dotted lines are ruled out. Dashed lines correspond to the situation where one of the two possible decays channels is still allowed.}
	\label{fig:Decays}
\end{figure*}

We show predictions for $\dnef$ generated from the different quark decays in Fig.~\ref{fig:Decays}. As already done before, we switch on only one coupling at a time. Both top decay channels yield the same $\dnef$ but they are subject to different bounds, and the same holds for bottom decays. In spite of the strong bounds on flavor violating couplings, given in Ref.~\cite{MartinCamalich:2020dfe} and reviewed in App.~\ref{app:bounds}, the signal is above $1\sigma$ in most of the cases.

Finally, we appreciate how the range of PQ breaking scales that could be detected through these hot axions in some frameworks overlap with that of cold axion dark matter. This is true for both scatterings and decays.  In particular, if the PQ symmetry is broken after inflation there is an additional contribution to axion dark matter from topological defects. Axions produced non-thermally through the decays of such defects are cold, and they are a viable dark matter candidate. Although there is a large theoretical uncertainty of this contribution, we claim for the benchmark value $f_a\gtrsim \cO(10^9) \GeV$ \cite{Gorghetto:2018myk,Gorghetto:2020qws} it could be possible to measure both hot and cold axions at the same time.

\subsection{UV Complete Models}
\label{sec:UV}

This section is devoted to the analysis of three specific models, the so-called DFSZ model~\cite{Zhitnitsky:1980tq,Dine:1981rt}, KSVZ model~\cite{Kim:1979if,Shifman:1979if} and the Minimal Flavor Violating Axion (MFVA) model~\cite{Arias-Aragon:2017eww}.

In the DFSZ case, the SM spectrum is supplemented by a second Higgs doublet and a scalar singlet. Both scalars and fermions transform under the Peccei-Quinn symmetry and, when the symmetry gets broken, an axion arises as a combination of the various Goldstone bosons. In particular, the axion couplings are flavor-blind and, using the notation of Eqs.~\eqref{eq:ReducedAxionCouplings1} and \eqref{eq:ReducedAxionCouplings2}, non-vanishing couplings with the top and bottom quarks are present, satisfying the following relation,
\be
c_t+c_b=\frac{1}{3}\,.
\ee

\begin{figure}
	\centering
	\includegraphics[height=0.32\textheight]{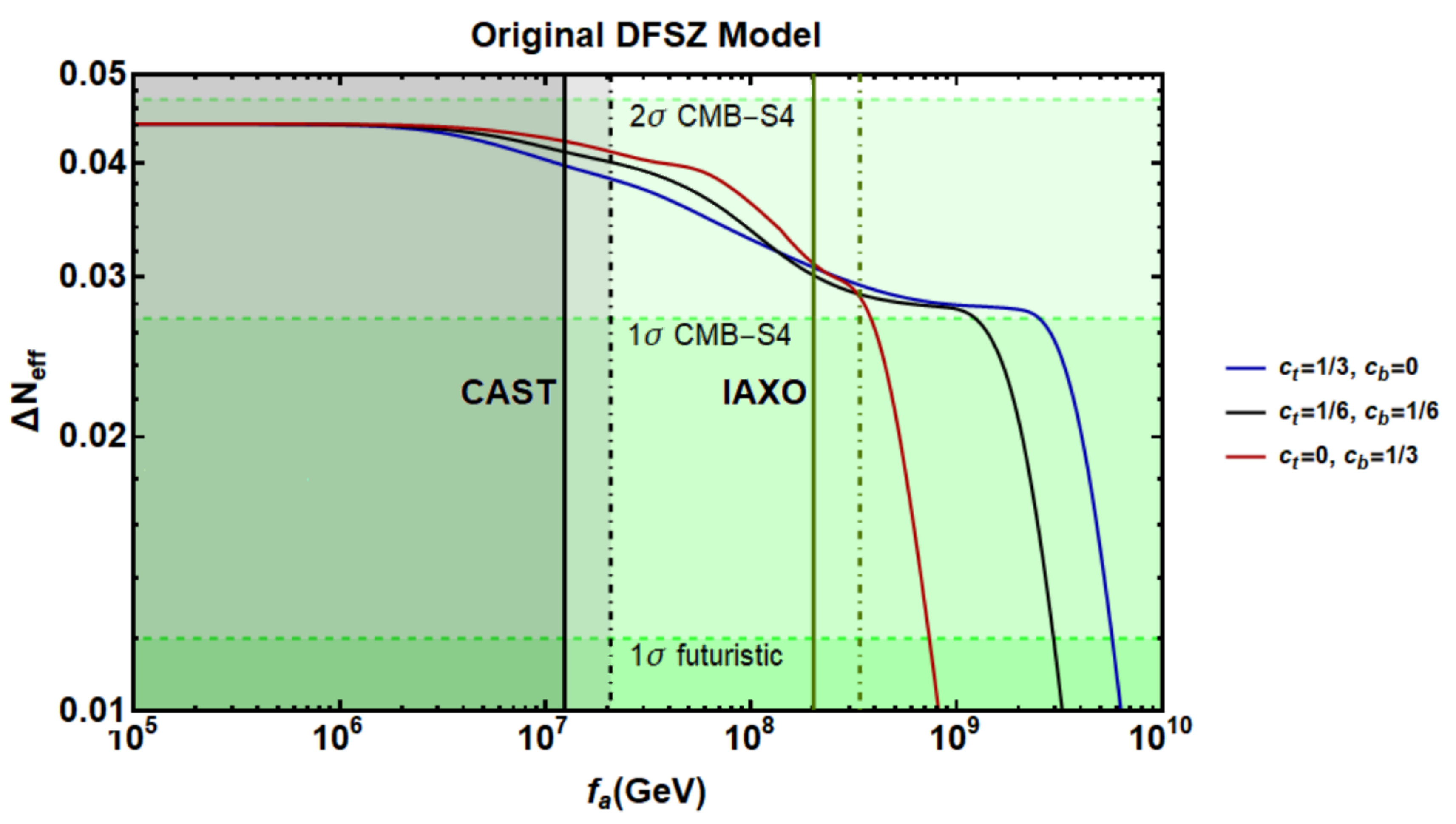}
	\caption{\em Total impact on $\dnef$ for a classic DFSZ axion. Three benchmarks couplings of the DFSZ axion to top and bottom quarks have been considered. The initial temperature here is set to $T_{I}=10^4 \,GeV$ and the initial axion abundance has been assumed to be zero. The CAST limit and IAXO prospect are shown: solid (dot-dashed) lines correspond to $c_{a\gamma\gamma}=8/3\ (2/3)$, when charged leptons couple to the same higgs doublet as the down-quarks (up-quarks) do.}
\label{figDFSZ}
\end{figure}

In the limit where all the scalar components are heavy, except for the would-be-longitudinal components of the gauge bosons, the physical $h$ and the axion, this model matches the general analyses performed in the previous sections. Notice that, dealing with a well-defined model, the $c_{a\gamma\gamma}$ coupling can be predicted in terms of the axion-fermion couplings: $c_{a\gamma\gamma}=2(4c_t+c_b+3c_\tau)$, where $c_\tau$ is the coupling with leptons and its defined in a similar way as $c_t$ and $c_b$ in Eqs.~\eqref{eq:ReducedAxionCouplings1} and \eqref{eq:ReducedAxionCouplings2}. Leptons can couple to the axion as the up-type quarks do or as the down-quarks do, and in general this leads to different values for the axion-photon coupling.

The contributions  to $\Delta \nef$ for the DSFZ model can be seen in Fig.~\ref{figDFSZ}. Three representative cases are considered, in order to cover the entire range of values for $c_b$ and $c_t$:  with $c_t=1/3$ and $c_b=0$ (in blue in the plot), with $c_t=0$ and $c_b=1/3$ (in red), and with $c_t=c_b=1/6$ (in black). For the CAST limit and IAXO prospect, the continuous line corresponds to the case with the smallest value for $c_{a\gamma\gamma}-1.92$, while the dot-dashed corresponds to the one with the most restrictive value of $c_{a\gamma\gamma}-1.92$. 

\begin{figure}
	\centering
	\includegraphics[height=0.32\textheight]{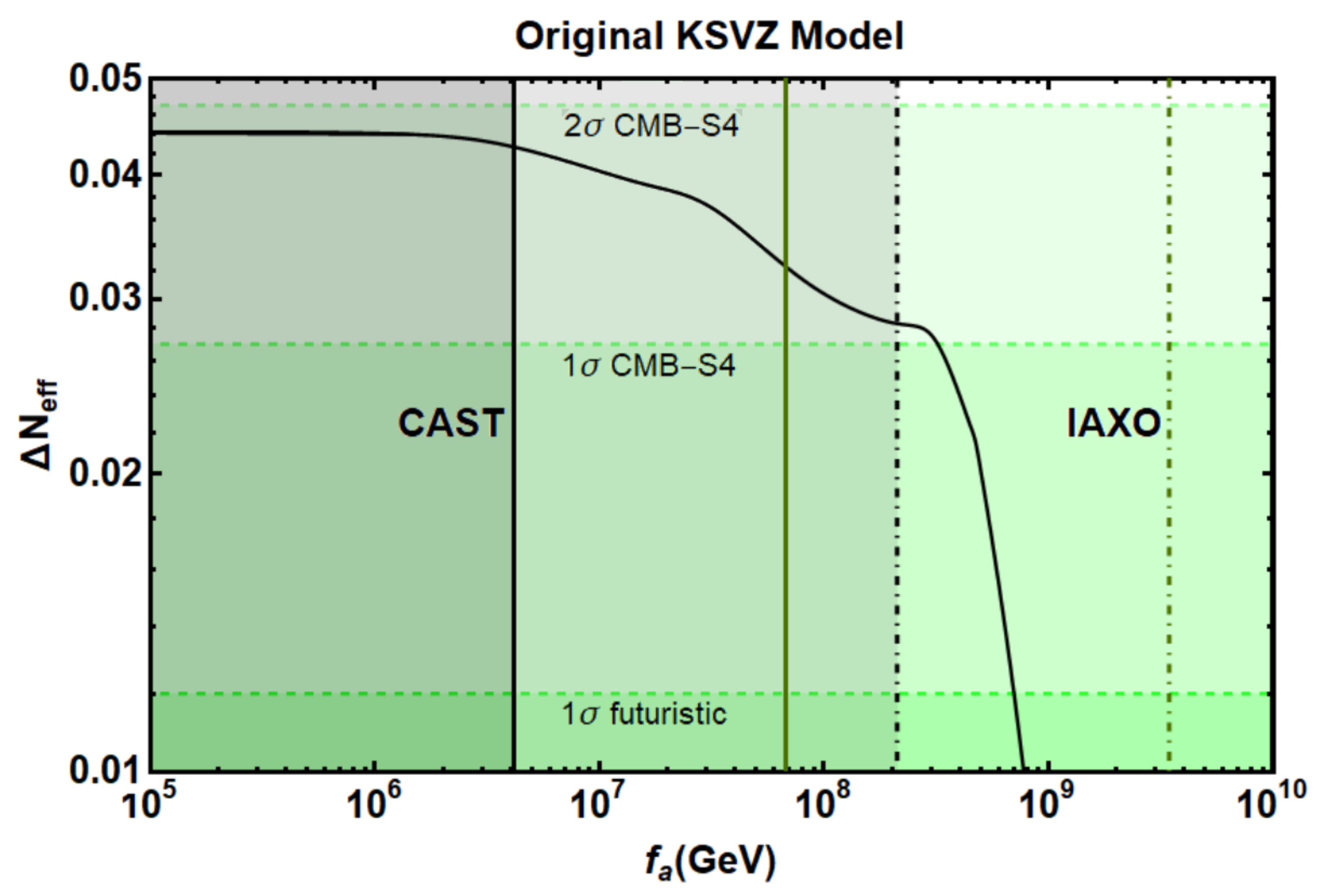}
	\caption{\em Total impact on $\dnef$ \, for a classic KSVZ axion. We set the initial temperature $T_{I}=10^4\, GeV$ and the initial axion abundance to zero. The CAST limit is shown as a shaded region, with solid lines corresponding to $c_{a\gamma\gamma}-1.92=-0.25$ and dot-dashed for $c_{a\gamma\gamma}-1.92=12.75$.}
	\label{figKSVZ}
\end{figure}

In the KSVZ model, the axion does not couple to the SM fermions at tree-level, but only to exotic quarks that enrich the SM fermionic spectrum. In this case, only an EW singlet scalar is added to the model and the axion arises as the Goldstone boson of this field, once the Peccei-Quinn symmetry gets broken. The only sizeable contributions to $\Delta \nef$ arise from the axion couplings to gluons, as axion couplings to SM fermions are induced only at 2-loops and therefore are strongly suppressed. Fig.~\ref{figKSVZ} shows the predictions for $\Delta \nef$ for this model. The range of axion-photon coupling considered here is $c_{a\gamma\gamma}-1.92\in \left[-0.25,\ 12.75\right]$, motivated by several possible UV completions of a KSVZ axion~\cite{DiLuzio:2017pfr}.

The MFVA model \cite{Arias-Aragon:2017eww}, instead, provides an effective description of the axion couplings with SM fields, once the flavor symmetry of the Minimal Flavor Violation framework \cite{Chivukula:1987py,DAmbrosio:2002vsn} is implemented in the Lagrangian. The axion couplings to fermions are universal within the same type of quarks and therefore are flavor conserving. Moreover, the axion coupling to up-type quarks is vanishing at leading order, and therefore the largest interactions are with the down-type quarks, and in particular with the bottom, due to Yukawa suppressions. This fact sensibly affects the results presented in the previous section, where the axion-top coupling was dominating all the contributions. In particular, the processes $b+\bar{b}\rightarrow g+a$ and $t+\bar{b}\rightarrow W^++a$ (proportional to the $ab\bar b$ coupling), that are proportional to the bottom quark Yukawa, were irrelevant when the axion coupled to the top, but now become crucial, as they are the only important contributions apart from the purely gluonic ones.

\begin{figure}
	\centering
	\includegraphics[height=0.32\textheight]{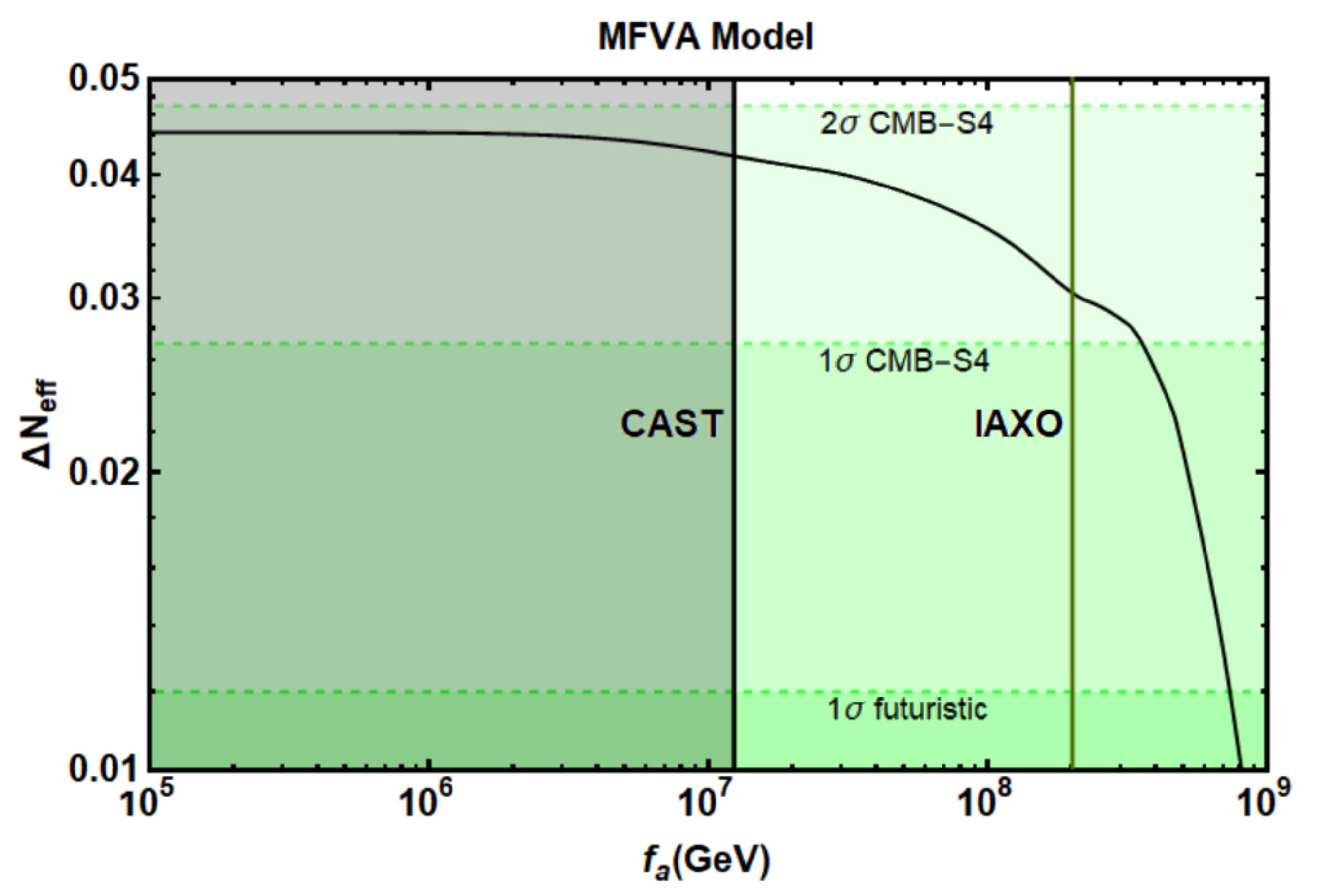}
	\caption{\em Total impact on $\dnef$ \, for the MFVA model. The initial temperature here is set to $T_{I}=10^4\GeV$ and the initial axion abundance is set to zero. The CAST limit and IAXO prospect are shown as a shaded region and a vertical green line, respectively.}
	\label{figMFVA}
\end{figure}

The coefficients describing axion couplings with bottoms $c_b$ and with photons $g_{a\gamma\gamma}$ acquire the following values in the MFVA model,
\be
c_b=\dfrac13\,,\qquad\qquad
g_{a\gamma\gamma}=\dfrac{\alpha_\text{em}}{2\pi}\dfrac{1}{f_a}\Bigg(\dfrac{8}{3}-1.92\Bigg)\,,
\ee
and the final result for $\Delta\nef$ is shown in Fig.~\eqref{figMFVA}.

As it can be seen, all models give the same contribution at low $f_a$. At high $f_a$, instead, the DFSZ model gives the largest abundances since it couples to all SM fermions already at tree level. For such a model one can reach a detectable axion abundance even in the range $f_a\approx 10^9-10^{10}$ GeV. If the PQ symmetry is broken after inflation and not restored afterwards, the abundance of cold axion dark matter receives a significant contribution from topological defects~\cite{Vilenkin:1982ks,Vachaspati:1984yi,Chang:1998tb,Hagmann:2000ja}. The detailed amount  from this source suffers a significant theoretical uncertainty~\cite{Gorghetto:2018myk,Gorghetto:2020qws}, but it is worth keeping in mind that in such a low $f_a$ region axion cold dark matter may coexist with detectable hot axions. Within the DFSZ framework, PQ symmetry in the post inflationary scenario has to be broken also explicitly to avoid the domain wall problem~\cite{Vilenkin:1981zs,Sikivie:1982qv,Ipser:1983db,Vilenkin:1984hy}.

Moreover, there is a window for $f_a$ between $10^7 \GeV$ and $2\times10^8 \GeV$ that can be explored by IAXO and is also above the $1\sigma$ level for the CMB-S4 experiments, where the models can be differentiated. This could imply an exciting opportunity to not only detect an effect of the axion, but also tell apart different invisible axion models.

When considering specific models with flavor violating axion couplings, like the Axiflavon~\cite{Calibbi:2016hwq} or Flaxion~\cite{Ema:2016ops}, they give a sizeable contribution to $\dnef\gtrsim0.01$ only for axion scales below $f_a\lesssim10^9\GeV$, a region which is largely excluded in those models due to the bound coming from the $K^+\rightarrow\pi^+a$ decay, being therefore irrelevant in this analysis.

\section{Conclusions}
\label{sec:conclusions}

The QCD axion is one of the best motivated candidates for physics beyond the SM. It provides an elegant dynamical solution to the strong CP problem and it is also a viable DM candidate. Recently, the community put forward a wealth of new ideas and techniques to detect such an elusive degree of freedom~\cite{Irastorza:2018dyq}. These experiments look for either virtual effects of a light pseudo-scalar mediator generating a new long range force or the axion DM wind within our Milky Way. 

There is an axion complementary probe within the reach of future experiments. Hot axions can be produced from scatterings or decays of thermal bath particles in the early universe, and they remain relativistic subsequently until the time of matter/radiation equality and recombination; this is true as long as $m_a\ll \cO\left(0.1\right)\eV$, as we consider in this work by neglecting the axion mass. They would manifest themselves in the CMB anisotropy spectrum as an additional radiation component, parameterized as the number of additional effective neutrinos $\dnef$.

In this work, we studied axions couplings to third generation quarks and we provided rigorous predictions for $\dnef$. We considered flavor conserving couplings, in which case production is controlled by binary collisions, and we also considered flavor violating couplings leading to axion production via two-body decays. We computed scattering cross sections and decay widths, and we obtained predictions for $\dnef$ after solving numerically the Boltzmann equation tracking the axion number density. Our predictions are smooth across the EWPT.

Our results can be found in Sec.~\ref{sec:results}. We studied both the model-independent contribution based on switching on an effective operator at a time as well as specific UV complete models. We found parameter space regions for all cases, typically with PQ breaking scale in the range $f_a \sim (10^9 - 10^{10}) \GeV$ for order one couplings to fermions, where the predicted signal is comparable to the forecasted $1\sigma$ sensitivity of CMB S4 experiments, and it could be detectable by more futuristic experiments. 

Finally, we point out two complementary signals. The values of the PQ breaking scale leading to an observable effect on $\dnef$ is consistent with axion cold dark matter. Furthermore, if there is no substantial hierarchy between the dimensionless axion couplings considered in this work and the associated one to photons then future helioscopes are also able to probe this parameter space region. The complementarity of these possible signals makes for a quite fascinating probe into the nature of the axion itself.


\paragraph{Acknowledgments.} F.A.A and L.M. acknowledge partial financial support by the Spanish MINECO through the Centro de excelencia Severo Ochoa Program under grant SEV-2016-0597, by the Spanish ``Agencia Estatal de Investigac\'ion''(AEI) and the EU ``Fondo Europeo de Desarrollo Regional'' (FEDER) through the projects FPA2016-78645-P and PID2019-108892RB-I00/AEI/10.13039/501100011033.
F.A.A, F.D. and L.M. acknowledge support from the European Union's Horizon 2020 research and innovation programme under the Marie Sk\l odowska-Curie grant agreement No 860881-HIDDeN.
The work of F.D. is supported by the grants: ``New Theoretical Tools for Axion Cosmology'' under the Supporting TAlent in ReSearch@University of Padova (STARS@UNIPD),  ``The Dark Universe: A Synergic Multi- messenger Approach'' number 2017X7X85K under the program PRIN 2017 funded by the Ministero dell'Istruzione, Universit\`a e della Ricerca (MIUR), ``New Theoretical Tools to Look at the Invisible Universe'' funded by the University of Padua. F.D. is also supported and by Istituto Nazionale di Fisica Nucleare (INFN) through the Theoretical Astroparticle Physics (TAsP) project.
 L.M. acknowledges partial financial support by the Spanish MINECO through the ``Ram\'on y Cajal'' programme (RYC-2015-17173). RZF acknowledges support by the Spanish Ministry MEC under grant  FPA 2017-88915-P and the Severo Ochoa excellence program of MINECO (SEV-2016- 0588).

\appendix

\section{Operator basis for axion couplings to quarks}
\label{app:basis}

In this appendix, we define the field basis for SM quarks that we employ in our analysis. The part of the SM Lagrangian needed for this discussion is the one containing Yukawa interactions. Focusing on quarks, the most generic set of Yukawa terms reads
\be
-\LL_Y = \ov{Q''_L}\,\widetilde{H}\,Y^u\,u''_R+\ov{Q''_L}\,H\,Y^d\,d''_R+\hc \ .
\label{YukawasChoiceDoublePrime}
\ee
The fields appearing in the operators above are: $SU(2)_L$ quark doublets $Q''_L$, $SU(2)_L$ quark singlets $u''_R$ and $d''_R$ and the $SU(2)_L$ Higgs doublet field $H$. Moreover, we define $\widetilde{H}\equiv i \sigma_2 (H^\dag)^T$ and $Y^{u,d}$ are generic $3\times 3$ diagonalizable matrices in flavor space. We save the symbol of unprimed fields for quark mass eigenstates defined later. 

We diagonalize the Yukawa matrices by performing bi-unitary transformations
\be
Y^u = U_{u_L} \, \widehat{Y}^u \, U_{u_R}^\dag\,,\qquad\qquad
Y^d = U_{d_L} \, \widehat{Y}^d \, U_{d_R}^\dag\,,
\ee
where the $U$ matrices are unitary and the hatted quantities are diagonal in flavor space. We introduce a new set of prime fields defined as follows
\be
Q''_L = U_{u_L}\,Q'_L\,,\qquad\qquad
u''_R = U_{u_R}\,u'_R\,,\qquad\qquad
d''_R = U_{d_R}\,d'_R\,,
\ee
The Yukawa Lagrangian in the new basis reads
\be
- \LL_Y = \ov{Q'_L}\,\widetilde{H}\,\widehat{Y}^u\,u'_R + \ov{Q'_L}\,H\,V_\text{CKM}\,\widehat{Y}^d\,d'_R+\hc \ ,
\label{YukawasChoicePrime}
\ee
with $V_\text{CKM} \equiv U^\dag_{u_L} \, U_{d_L}$ the CKM matrix. In our study, we always specify axion couplings in the primed field basis for quarks with Yukawa interactions as in Eq.~\eqref{YukawasChoicePrime}.

Finally, we identify the quark mass eigenstates, which we denote with unprimed fields, and their relation to the primed fields. First, we identify the components of the quark doublet $Q'_L = (u'_L \; d'_L)$. Once the Higgs gets a vacuum expectation value (vev), we identify mass eigenstates by redefining the left-handed down quarks
\be
u'_L=u_L\,,\qquad\qquad
d'_L=V_{CKM}\,d_L\,,\qquad\qquad
u'_R=u_R\,,\qquad\qquad
d'_R=d_R\,,
\label{eq:massbasisrotation}
\ee
Flavor eigenstates $u_i'$ coincide with the mass eigenstates $u_i$, and the $b'$ quark, the only down-quark we are interested in, almost coincides with the $b$ quark up to CKM corrections of order $\cO(0.05)$. In contrast to gauge interactions in the primed basis, which are still flavor diagonal, the CKM matrix appears in the fermion charged current once we switch to mass eigenstates
\be
J_\mu^-=\ov{u_L} \, \gamma_\mu \, V_\text{CKM}\, d_L \ .  
\label{CKMcurrent}
\ee

\section{Cross sections below the EWPT}
\label{app:XSbelow}

We provide analytical cross sections for the processes listed in Tab.~\ref{tab:below}, and we begin with the first block where the two particles in the initial state are fermions. A quark can find its own antiparticle and annihilate to final states containing one axion particle. If the final state is the gluon we have quark-antiquark annihilations to gluon and axion with a cross section
\begin{equation}
\sigma_{q\bar{q} \rightarrow ga} = \frac{ c_q^2 g_s^2 m_{q}^2}{9\pi  f_a^2 \left(s-4 m_{q}^2\right)}\tanh ^{-1}\left(\sqrt{1-\frac{4 m_{q}^2}{s}}\right) \ ,
\label{eq:qqga}
\end{equation}
where $g_s$ is the strong coupling constant and $s$ is the usual Mandelstam variable denoting the (squared of the) energy in the center of mass frame. Here and below, we denote with the letter $q = \{t, b\}$ a generic third generation quark when it is possible to provide a single expression valid for both cases. If the other SM particle in the final state is the Higgs boson we have
\be
\sigma_{q\bar{q} \rightarrow h a}^\downarrow=
\scriptstyle\frac{c_q^2 y_q^2 \left(s-m_h^2\right)}{64 \pi  s f_a^2 \left(s-4 m_q^2\right)} \left(\sqrt{s \left(s-4 m_q^2\right)}-4 m_q^2 \tanh ^{-1}\left(\sqrt{1-\frac{4 m_q^2}{s}}\right)\right) \ .
\label{eq:qqha}
\ee
where the symbol ``$\downarrow$'' indicates that cross sections are calculated below the EWPT. Likewise, quarks can annihilate with their own antiquarks leading to an axion final state together with the $Z$ boson with cross sections
\be\label{eq:ttZa}
\begin{aligned}
\sigma_{t \bar{t} \rightarrow Z a}=&
\scriptstyle\frac{c_t^2 g_W^2 m_t^2 \sqrt{\frac{1}{s-4 m_t^2}} \left(s-m_Z^2\right)}{1152 \pi  s^{3/2} f_a^2 m_W^2 m_Z^2}\left(4 \sqrt{\frac{s}{s-4 m_t^2}} \tanh ^{-1}\left(\sqrt{1-\frac{4 m_t^2}{s}}\right) \left(-m_Z^2 \left(9 m_t^2+40 m_W^2\right)+32 m_W^4+17 m_Z^4\right)+9 m_Z^2
   \left(s-2 m_Z^2\right)\right)\,,
\end{aligned}
\ee
\be \label{eq:bbZa}
\begin{aligned}
\sigma_{b \bar{b} \rightarrow Z a}=&
\scriptstyle\frac{c_b^2 g_W^2 m_b^2 \sqrt{\frac{1}{s-4 m_b^2}} \left(s-m_Z^2\right)}{1152 \pi  s^{3/2} f_a^2 m_W^2 m_Z^2}\left(4 \sqrt{\frac{s}{s-4 m_b^2}} \tanh ^{-1}\left(\sqrt{1-\frac{4 m_b^2}{s}}\right) \left(-m_Z^2 \left(9 m_b^2+4 m_W^2\right)+8 m_W^4+5 m_Z^4\right)+9 m_Z^2
   \left(s-2 m_Z^2\right)\right)\,,
\end{aligned}
\ee
Finally, to complete the first block of the table, we can have a top quark and a bottom antiquark as well as the CP conjugate system annihilating to a final state with an axion and a $W$ boson with cross section
\begin{align}\label{eq:tbWa}
\sigma_{t \bar{b} \rightarrow W^+ a}=&
\scriptstyle\frac{g_W^2 \left(s-m_W^2\right)}{128 \pi  s f_a^2 m_W^2 \left(\left(-m_b^2+m_t^2+s\right)^2-4 s m_t^2\right)}\Bigg(\left(c_b^2m_b^2+c_t^2m_t^2\right)\left(s-2 m_W^2\right) \sqrt{-2 m_b^2 \left(m_t^2+s\right)+m_b^4+\left(m_t^2-s\right)^2}+\nn\\
&\scriptstyle-2 c_t^2m_t^2s \left(m_b^2-m_t^2+2 m_W^2\right) \coth^{-1}\left(\frac{m_b^2-m_t^2-s}{\sqrt{-2 m_b^2 \left(m_t^2+s\right)+m_b^4+\left(m_t^2-s\right)^2}}\right)+\nn\\
&\scriptstyle+2c_bm_b^2s\bigg(2c_tm_t^2\coth^{-1}\left(\frac{m_b^2+m_t^2-s}{\sqrt{-2 m_b^2 \left(m_t^2+s\right)+m_b^4+\left(m_t^2-s\right)^2}}\right)+\nn\\
&\scriptstyle+c_b\left(m_t^2-m_b^2+2m_W^2\right)\coth^{-1}\left(\frac{m_b^2-m_t^2+s}{\sqrt{-2 m_b^2 \left(m_t^2+s\right)+m_b^4+\left(m_t^2-s\right)^2}}\right)\bigg)\Bigg)\,,
\end{align}

We switch to the second block of Tab.~\ref{tab:below} and we consider when there is just one fermion in the initial and final states. For a gluon in the initial state we find
\begin{equation}
\sigma_{q g \rightarrow q a} = \frac{c_q^2 g_s^2 m_q^2}{192\pi  f_a^2 s^2 \left(s-m_q^2\right)} \left[2 s^2 \log \left(\frac{s}{m_q^2}\right)+4 s m_q^2-m_q^4-3 s^2\right] \ .
\label{eq:qgqa}
\end{equation}
For quark/Higgs boson scattering we have
\be
\begin{aligned}
\sigma_{q h \rightarrow q a}^\downarrow=&
\scriptstyle\frac{c_q^2 y_q^2 \left(s-m_q^2\right)}{64 \pi  s f_a^2 \left(\left(-m_h^2+m_q^2+s\right)^2-4 s m_q^2\right)} \Bigg[\left(-m_h^2+m_q^2+s\right) \sqrt{-2 m_h^2 \left(m_q^2+s\right)+m_h^4+\left(m_q^2-s\right)^2}+\\
&\scriptstyle-2 s m_q^2 \log\left(-\frac{\sqrt{-2 m_h^2 \left(m_q^2+s\right)+m_h^4+\left(m_q^2-s\right)^2}-m_h^2+m_q^2+s}{\sqrt{-2 m_h^2\left(m_q^2+s\right)+m_h^4+\left(m_q^2-s\right)^2}+m_h^2-m_q^2-s}\right)\Bigg] \ ,
\end{aligned}
\label{eq:qhqa}
\ee
whereas for the case of a $Z$ boson we find
\be\label{eq:tZta}
\begin{aligned}
\sigma_{t Z \rightarrow t a}=&
\scriptstyle\frac{c_t^2 g_W^2 m_t^2 \left(s-m_t^2\right)}{3456 \pi  s^2 f_a^2 m_W^2 m_Z^2 \sqrt{\left(m_t^2-m_Z^2+s\right)^2-4 s m_t^2}}\times\\
&\scriptstyle \times\Bigg(\frac{2 s^2 \left(m_Z^2 \left(9 m_t^2+40 m_W^2\right)-32 m_W^4-17 m_Z^4\right) \log \left(\frac{-\sqrt{-2 m_t^2 \left(m_Z^2+s\right)+m_t^4+\left(m_Z^2-s\right)^2}+m_t^2-m_Z^2+s}{\sqrt{-2 m_t^2\left(m_Z^2+s\right)+m_t^4+\left(m_Z^2-s\right)^2}+m_t^2-m_Z^2+s}\right)}{\sqrt{-2 m_t^2 \left(m_Z^2+s\right)+m_t^4+\left(m_Z^2-s\right)^2}}+\\
&\scriptstyle+3 s \left(m_Z^2 \left(3 m_t^2+40 m_W^2\right)-32 m_W^4-8 m_Z^4\right)+\left(m_t-m_Z\right) \left(m_t+m_Z\right) \left(-40 m_W^2 m_Z^2+32 m_W^4+17
   m_Z^4\right)+9 s^2 m_Z^2\Bigg)\,.
\end{aligned}
\ee
\be \label{eq:bZba}
\begin{aligned}
\sigma_{b Z \rightarrow b a}=&
\scriptstyle\frac{c_b^2 g_W^2 m_b^2 \left(s-m_b^2\right)}{3456 \pi  s^2 f_a^2 m_W^2 m_Z^2 \sqrt{\left(m_b^2-m_Z^2+s\right)^2-4 s m_b^2}}\times\\
&\scriptstyle \times\Bigg(\frac{2 s^2 \left(m_Z^2 \left(9 m_b^2+4 m_W^2\right)-8 m_W^4-5 m_Z^4\right) \log \left(\frac{-\sqrt{-2 m_b^2 \left(m_Z^2+s\right)+m_b^4+\left(m_Z^2-s\right)^2}+m_b^2-m_Z^2+s}{\sqrt{-2 m_b^2\left(m_Z^2+s\right)+m_b^4+\left(m_Z^2-s\right)^2}+m_b^2-m_Z^2+s}\right)}{\sqrt{-2 m_b^2 \left(m_Z^2+s\right)+m_b^4+\left(m_Z^2-s\right)^2}}+\\
&\scriptstyle+3 s \left(m_Z^2 \left(3 m_b^2+4 m_W^2\right)-8 m_W^4+4 m_Z^4\right)+\left(m_b-m_Z\right) \left(m_b+m_Z\right) \left(-4 m_W^2 m_Z^2+8 m_W^4+5 m_Z^4\right)+9 s^2 m_Z^2\Bigg)\,.
\end{aligned}
\ee
Finally, if the initial state quark annihilate with a $W$ boson we have the cross sections
\begin{align} \label{eq:tWba}
\sigma_{tW^-\rightarrow ba}=&
\scriptstyle\frac{g_W^2 \left(s-m_b^2\right)}{384 \pi  s^2 f_a^2 m_W^2 \left(\left(m_t^2-m_W^2+s\right)^2-4 s m_t^2\right)}\Bigg(\sqrt{-2 m_t^2 \left(m_W^2+s\right)+\left(m_W^2-s\right)^2+m_t^4}\Big(2c_bc_tm_b^2m_t^2\left(3s+m_W^2-m_t^2\right)+\nn\\
&\scriptstyle+c_b^2m_b^2\left(m_t^4-2m_W^4+m_t^2\left(m_W^2-2s\right)+m_W^2s+s^2\right)+c_t^2m_t^2\left(m_b^2\left(m_t^2-m_W^2-3s\right)+s\left(m_t^2-m_W^2+s\right)\right)\Big)+\nn\\
&\scriptstyle+2s^2c_tm_t^2\left(2c_bm_b^2+c_t\left(m_t^2-m_b^2-2m_W^2\right)\right) \log\left(\frac{m_t^2-m_W^2+s-\sqrt{-2 m_t^2\left(m_W^2+s\right)+\left(m_W^2-s\right)^2+m_t^4}}{m_t^2-m_W^2+s+\sqrt{-2 m_t^2\left(m_W^2+s\right)+\left(m_W^2-s\right)^2+m_t^4}}\right)\Bigg)\,,
\end{align}

\begin{align} \label{eq:bWta}
\sigma_{bW^+\rightarrow ta}=&
\scriptstyle\frac{g_W^2 \left(s-m_t^2\right)}{384 \pi  s^2 f_a^2 m_W^2 \left(\left(m_b^2-m_W^2+s\right)^2-4 s m_b^2\right)}\Bigg(\sqrt{-2 m_b^2 \left(m_W^2+s\right)+\left(m_W^2-s\right)^2+m_b^4}\Big(2c_tc_bm_t^2m_b^2\left(3s+m_W^2-m_b^2\right)+\nn\\
&\scriptstyle+c_t^2m_t^2\left(m_b^4-2m_W^4+m_b^2\left(m_W^2-2s\right)+m_W^2s+s^2\right)+c_b^2m_b^2\left(m_t^2\left(m_b^2-m_W^2-3s\right)+s\left(m_b^2-m_W^2+s\right)\right)\Big)+\nn\\
&\scriptstyle+2s^2c_bm_b^2\left(2c_tm_t^2+c_b\left(m_b^2-m_t^2-2m_W^2\right)\right) \log\left(\frac{m_b^2-m_W^2+s-\sqrt{-2 m_b^2\left(m_W^2+s\right)+\left(m_W^2-s\right)^2+m_b^4}}{m_b^2-m_W^2+s+\sqrt{-2 m_b^2\left(m_W^2+s\right)+\left(m_W^2-s\right)^2+m_b^4}}\right)\Bigg)\,,
\end{align}

\section{Bounds on axion couplings}
\label{app:bounds}

In this Appendix, we collect astrophysical, cosmological and terrestrial experimental bounds on axion interactions with SM particles given in Sec.~\ref{sec:Lagrangian}.

\begin{description}
\item[Coupling with photons.] The effective coupling to photons, as defined in Eq.~\eqref{caggcagammagammadef}, must satisfy the following constraints~\cite{Millea:2015qra,Jaeckel:2015jla,Bauer:2017ris,Anastassopoulos:2017ftl}:
\be
\begin{aligned}
|g_{a\gamma\gamma}|\lesssim&\,7\times 10^{-11}\GeV^{-1}
&&\text{for}\quad m_a\lesssim10 \meV
\\
|g_{a\gamma\gamma}|\lesssim&\, 10^{-10}\GeV^{-1}
&&\text{for}\quad 10 \meV\lesssim m_a\lesssim1 \eV
\\
|g_{a\gamma\gamma}|\ll&\, 10^{-12}\GeV^{-1}
&&\text{for}\quad 10 \eV\lesssim m_a\lesssim0.1\GeV
\\
|g_{a\gamma\gamma}|\lesssim&\, 10^{-3}\GeV^{-1}
&&\text{for}\quad 0.1 \GeV\lesssim m_a\lesssim1\TeV\,.
\end{aligned}
\label{ConstraintsAxion2gamma}
\ee
For masses larger than the TeV, no constraint is present on these couplings. These bounds can be translated in terms of $f_a$ once a specific value of $c_{a\gamma\gamma}$ is taken.

\item[Axion flavor conserving couplings to third generation quarks.] Stellar cooling data imply bounds on axion couplings to top and bottom quarks~\cite{Feng:1997tn}. In general, this constraint applies on the effective axion coupling with electrons, which is the sum between the tree-level coupling with electrons and the loop-induced contributions proportional to the axion couplings with any other fermion. Under the assumption that only one coupling is non-vanishing at a time, and in particular the tree level coupling with electrons is zero, then
\be
\dfrac{f_a}{c_t}\gtrsim1.2\times10^{9}\GeV\qquad\qquad
\dfrac{f_a}{c_b}\gtrsim6.1\times10^{5}\GeV\,,
\label{eq:boundsFC}
\ee
for axion masses in the range $m_a\lesssim10\keV$.

\item[Axion couplings to nucleons.] Neutron star and Supernova SN1987A cooling data provide bounds on axion coupling with neutrons and nuclei. The physical process consists in the neutron or nucleus bremsstrahlung, respectively, and the corresponding bounds read
\be
\dfrac{f_a}{c_{an}}>1.21\times 10^9\GeV\text{~\cite{Keller:2012yr,Sedrakian:2015krq}}\qquad\qquad
\dfrac{f_a}{\sqrt{c_{ap}^2+c_{an}^2}}>1.67\times 10^9\GeV\text{~\cite{Fischer:2016cyd,Giannotti:2017hny}}\,,
\ee
where $c_{an}$ and $c_{ap}$ stand for the effective coupling of axion to neutrons and protons and are expressed in terms of the axion-quark couplings as follows:
\be
\begin{gathered}
\textstyle c_{an} = -0.02+0.88c_{d}-0.39c_{u}-0.038c_{s}-0.012c_{c}-0.009c_{b}-0.0035c_{t},\\
c_{ap} = -0.47+0.88c_{u}-0.39c_{d}-0.038c_{s}-0.012c_{c}-0.009c_{b}-0.0035c_{t}.
\end{gathered}
\ee
The constant terms refer to the axion coupling to gluons, while the others to the corresponding axion-fermion couplings. These bounds are rather strong, but should be taken with caution: from one side they are model dependent and from the other hold under the current knowledge of the complicated Supernova physics and neutron stars. If one considers only the couplings $c_t$ and $c_b$, as we do in our work, these bounds are sub-dominant with respect to the ones in Eq.~\eqref{eq:boundsFC}.

\item[Axion flavor violating couplings to third generation quarks.] Flavor violating coupling are strongly constrained from processes like rare decays or meson oscillations. An example of these processes are $B^+\rightarrow \pi^+a$ decay and $B^0-\bar{B}^0$ oscillations, from where a bound can be obtained on the vector and axial axion coupling, respectively, to bottom and down quarks \cite{MartinCamalich:2020dfe,Aubert:2004ws,UTfit}:
\begin{equation}
\frac{f_a}{c_{bd}^V}>1.1\times10^8\GeV\qquad\qquad \frac{f_a}{c_{bd}^A}>2.6\times10^6\GeV.
\end{equation}
Analogously, from the processes $B^{+,0}\rightarrow{K}^{+,0}a$ and $B^{+,0}\rightarrow {K^*}^{+,0}a$ bounds on the vector and axial couplings to bottom and strange quarks can be obtained~\cite{Lees:2013kla}:
\begin{equation}
\frac{f_a}{c_{bs}^V}>3.3\times10^8\GeV\qquad\qquad \frac{f_a}{c_{bs}^A}>1.3\times10^8\GeV.
\end{equation}
Bounds of flavor violating couplings involving the top quark are obtained in the same fashion as for the flavor conserving ones: considering the contribution at one loop to the process $K^+\rightarrow\pi^+a$ of a top-up and top-charm coupling it is possible to extract the following bounds~\cite{Adler:2008zza}:
\begin{equation}
\frac{f_a}{c_{tu}}>3\times10^8\GeV\qquad\qquad \frac{f_a}{c_{tc}}>7\times10^8\GeV.
\end{equation}

\end{description}

\section{Approaching the QCDPT}

\begin{figure}
	\centering
	\includegraphics[width=0.45 \linewidth]{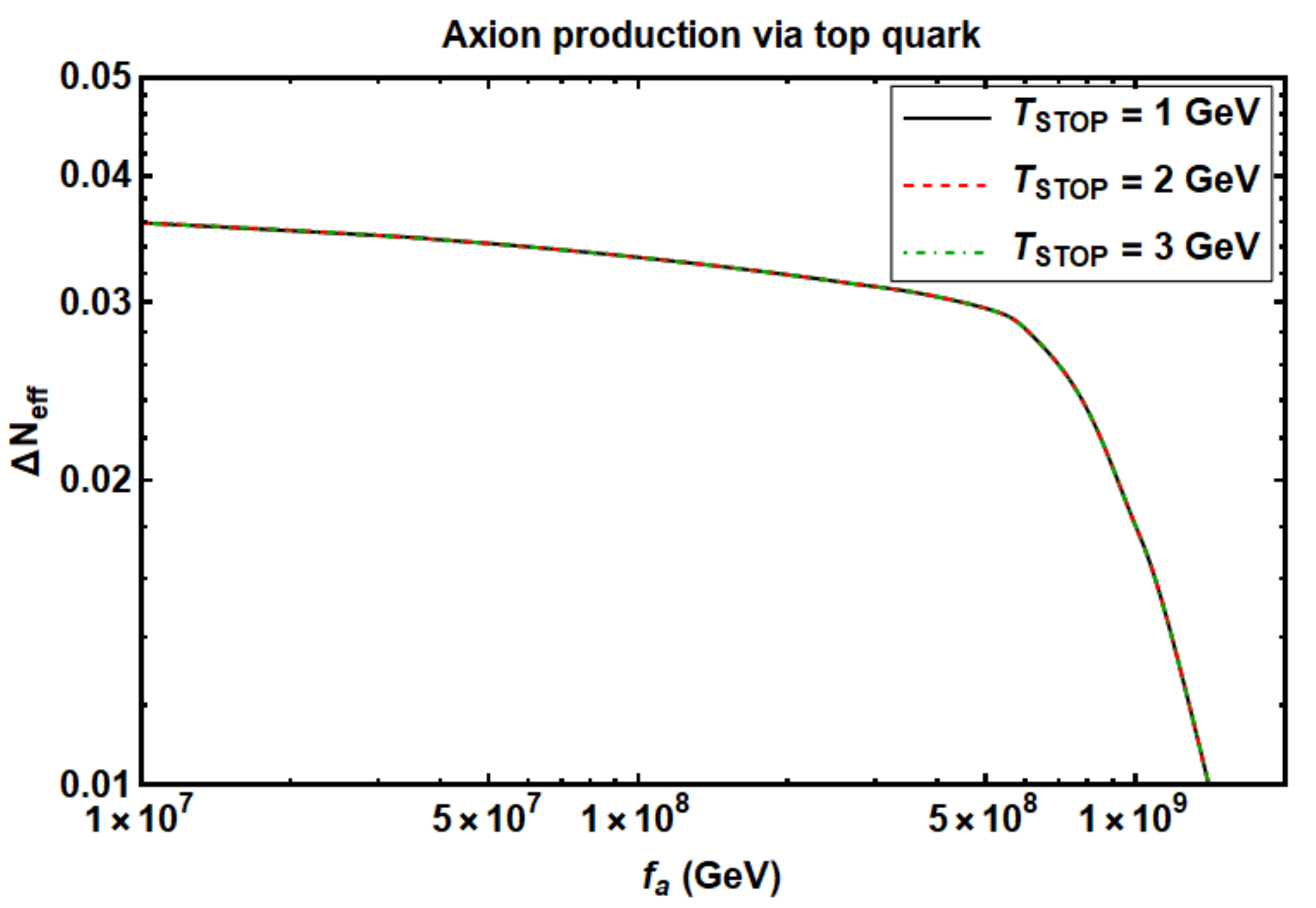} $\qquad$
	\includegraphics[width=0.45 \linewidth]{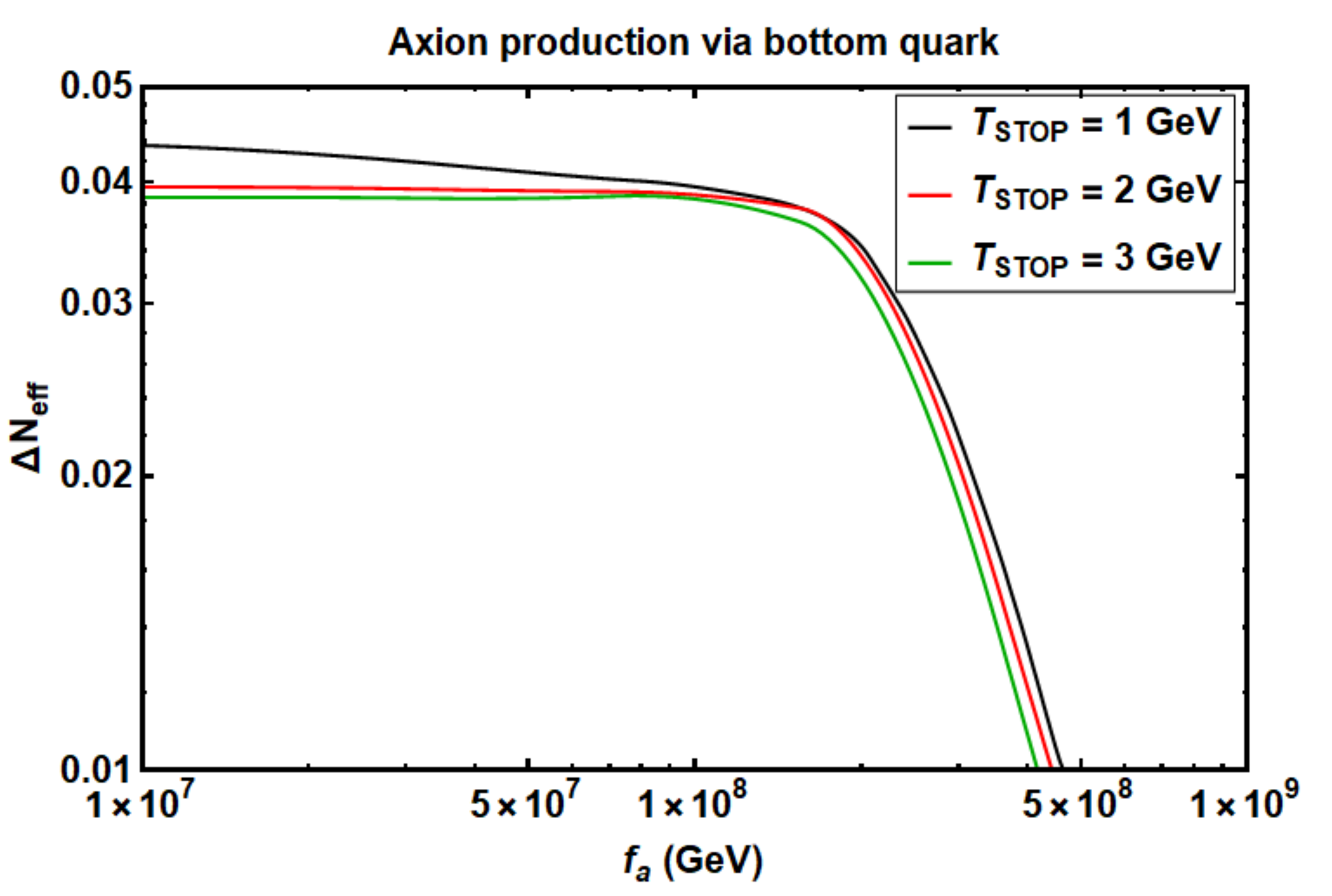}
	\caption{\em Sensitivity of the $\Delta N_{eff}$ prediction on the lowest temperature $T_{\rm STOP}$ reached by our Boltzmann equation integration. We choose values of $T_{\rm STOP}$ close to the QCDPT, and we show results for production via top (left panel) and bottom (right panel) scattering.}
	\label{figTSTOP}
\end{figure}

Ideally, we should integrate the Boltzmann equation tracking the axion number density all the way down to very low temperatures in order to predict $\Delta \nef$. We have seen why this is not necessary because the axion comoving density reaches an asymptotic value once SM quarks participating in the production starts feeling the Maxwell-Boltzmann suppression. So it is enough to stop our Boltzmann equation integration at some IR temperature cutoff that we denote $T_{\rm STOP}$.

The value of the needed $T_{\rm STOP}$ could be dangerous if it is too low. Our analysis is based on perturbative calculations for scattering cross sections and on treating the primordial bath as a gas of weakly-coupled quarks and gluons in thermal equilibrium. This setup is certainly valid at high temperatures around the EWPT, and it loses its validity as we approach the QCDPT. In this appendix, we investigate how robust is our predictions for $\Delta \nef$ considering this potential issue.

We show in Fig.~\ref{figTSTOP} the prediction for $\Delta \nef$ as a function of $f_a$ for axion production via top quark (left panel) and bottom quark (right panel) scatterings. In each panel, we report our prediction for the different values $T_{\rm STOP} = \left\{1, 2, 3\right\} \, {\rm GeV}$ close to the QCDPT. The result for the top is absolutely stable, and this is not surprising since the top mass is much larger than the typical temperatures around the QCDPT. On the contrary, production via bottom scattering presents some dependence on this temperature and decreasing it leads to slightly higher $\dnef$. However, such a dependence on $T_{\rm STOP}$ is noticeable mostly in the region of very low PQ breaking scales ruled out by experiments. Thus, our results are robust. And, in any case, they could be interpreted as a lower bound on the expected effect on $\dnef$ that still ensures perturbativity in the computations.


\bibliographystyle{JHEP}
\bibliography{refs}

\providecommand{\href}[2]{#2}\begingroup\raggedright\begin{thebibliography}{10}

\bibitem{Baker:2006ts}
C.A.~Baker et~al., \emph{{An Improved Experimental Limit on the Electric Dipole
  Moment of the Neutron}},
  \href{https://doi.org/10.1103/PhysRevLett.97.131801}{\emph{Phys. Rev. Lett.}
  {\bfseries 97} (2006) 131801}
  [\href{https://arxiv.org/abs/hep-ex/0602020}{{\ttfamily hep-ex/0602020}}].

\bibitem{Afach:2015sja}
J.M.~Pendlebury et~al., \emph{{Revised Experimental Upper Limit on the Electric
  Dipole Moment of the Neutron}},
  \href{https://doi.org/10.1103/PhysRevD.92.092003}{\emph{Phys. Rev.}
  {\bfseries D92} (2015) 092003}
  [\href{https://arxiv.org/abs/1509.04411}{{\ttfamily 1509.04411}}].

\bibitem{Peccei:1977hh}
R.D.~Peccei and H.R.~Quinn, \emph{{CP Conservation in the Presence of
  Instantons}}, \href{https://doi.org/10.1103/PhysRevLett.38.1440}{\emph{Phys.
  Rev. Lett.} {\bfseries 38} (1977) 1440}.

\bibitem{Peccei:1977ur}
R.D.~Peccei and H.R.~Quinn, \emph{{Constraints Imposed by CP Conservation in
  the Presence of Instantons}},
  \href{https://doi.org/10.1103/PhysRevD.16.1791}{\emph{Phys. Rev.} {\bfseries
  D16} (1977) 1791}.

\bibitem{Wilczek:1977pj}
F.~Wilczek, \emph{{Problem of Strong P and T Invariance in the Presence of
  Instantons}}, \href{https://doi.org/10.1103/PhysRevLett.40.279}{\emph{Phys.
  Rev. Lett.} {\bfseries 40} (1978) 279}.

\bibitem{Weinberg:1977ma}
S.~Weinberg, \emph{{A New Light Boson?}},
  \href{https://doi.org/10.1103/PhysRevLett.40.223}{\emph{Phys. Rev. Lett.}
  {\bfseries 40} (1978) 223}.

\bibitem{Vafa:1984xg}
C.~Vafa and E.~Witten, \emph{{Parity Conservation in QCD}},
  \href{https://doi.org/10.1103/PhysRevLett.53.535}{\emph{Phys. Rev. Lett.}
  {\bfseries 53} (1984) 535}.

\bibitem{Bardeen:1978nq}
W.A.~Bardeen, S.H.H.~Tye and J.A.M.~Vermaseren, \emph{{Phenomenology of the New
  Light Higgs Boson Search}},
  \href{https://doi.org/10.1016/0370-2693(78)90859-6}{\emph{Phys. Lett.}
  {\bfseries 76B} (1978) 580}.

\bibitem{diCortona:2015ldu}
G.~Grilli~di Cortona, E.~Hardy, J.~Pardo~Vega and G.~Villadoro, \emph{{The QCD
  Axion, Precisely}},
  \href{https://doi.org/10.1007/JHEP01(2016)034}{\emph{JHEP} {\bfseries 01}
  (2016) 034} [\href{https://arxiv.org/abs/1511.02867}{{\ttfamily
  1511.02867}}].

\bibitem{Anastassopoulos:2017ftl}
{\scshape CAST} collaboration, \emph{{New Cast Limit on the Axion-Photon
  Interaction}}, \href{https://doi.org/10.1038/nphys4109}{\emph{Nature Phys.}
  {\bfseries 13} (2017) 584}
  [\href{https://arxiv.org/abs/1705.02290}{{\ttfamily 1705.02290}}].

\bibitem{Jaeckel:2015jla}
J.~Jaeckel and M.~Spannowsky, \emph{{Probing MeV to 90 GeV Axion-Like Particles
  with Lep and Lhc}},
  \href{https://doi.org/10.1016/j.physletb.2015.12.037}{\emph{Phys. Lett.}
  {\bfseries B753} (2016) 482}
  [\href{https://arxiv.org/abs/1509.00476}{{\ttfamily 1509.00476}}].

\bibitem{Brivio:2017ije}
I.~Brivio, M.B.~Gavela, L.~Merlo, K.~Mimasu, J.M.~No, R.~del Rey et~al.,
  \emph{{Alps Effective Field Theory and Collider Signatures}},
  \href{https://doi.org/10.1140/epjc/s10052-017-5111-3}{\emph{Eur. Phys. J.}
  {\bfseries C77} (2017) 572}
  [\href{https://arxiv.org/abs/1701.05379}{{\ttfamily 1701.05379}}].

\bibitem{Bauer:2017ris}
M.~Bauer, M.~Neubert and A.~Thamm, \emph{{Collider Probes of Axion-Like
  Particles}}, \href{https://doi.org/10.1007/JHEP12(2017)044}{\emph{JHEP}
  {\bfseries 12} (2017) 044}
  [\href{https://arxiv.org/abs/1708.00443}{{\ttfamily 1708.00443}}].

\bibitem{Irastorza:2018dyq}
I.G.~Irastorza and J.~Redondo, \emph{{New experimental approaches in the search
  for axion-like particles}},
  \href{https://doi.org/10.1016/j.ppnp.2018.05.003}{\emph{Prog. Part. Nucl.
  Phys.} {\bfseries 102} (2018) 89}
  [\href{https://arxiv.org/abs/1801.08127}{{\ttfamily 1801.08127}}].

\bibitem{Kim:1979if}
J.E.~Kim, \emph{{Weak Interaction Singlet and Strong CP Invariance}},
  \href{https://doi.org/10.1103/PhysRevLett.43.103}{\emph{Phys. Rev. Lett.}
  {\bfseries 43} (1979) 103}.

\bibitem{Shifman:1979if}
M.A.~Shifman, A.I.~Vainshtein and V.I.~Zakharov, \emph{{Can Confinement Ensure
  Natural CP Invariance of Strong Interactions?}},
  \href{https://doi.org/10.1016/0550-3213(80)90209-6}{\emph{Nucl. Phys.}
  {\bfseries B166} (1980) 493}.

\bibitem{Zhitnitsky:1980tq}
A.R.~Zhitnitsky, \emph{{On Possible Suppression of the Axion Hadron
  Interactions. (In Russian)}}, {\emph{Sov. J. Nucl. Phys.} {\bfseries 31}
  (1980) 260}.

\bibitem{Dine:1981rt}
M.~Dine, W.~FisCHLer and M.~Srednicki, \emph{{A Simple Solution to the Strong
  CP Problem with a Harmless Axion}},
  \href{https://doi.org/10.1016/0370-2693(81)90590-6}{\emph{Phys. Lett.}
  {\bfseries B104} (1981) 199}.

\bibitem{Marsh:2015xka}
D.J.E.~Marsh, \emph{{Axion Cosmology}},
  \href{https://doi.org/10.1016/j.physrep.2016.06.005}{\emph{Phys. Rept.}
  {\bfseries 643} (2016) 1} [\href{https://arxiv.org/abs/1510.07633}{{\ttfamily
  1510.07633}}].

\bibitem{Turner:1986tb}
M.S.~Turner, \emph{{Thermal Production of Not SO Invisible Axions in the Early
  Universe}}, \href{https://doi.org/10.1103/PhysRevLett.59.2489,
  10.1103/PhysRevLett.60.1101.3}{\emph{Phys. Rev. Lett.} {\bfseries 59} (1987)
  2489}.

\bibitem{Masso:2002np}
E.~Masso, F.~Rota and G.~Zsembinszki, \emph{{On axion thermalization in the
  early universe}},
  \href{https://doi.org/10.1103/PhysRevD.66.023004}{\emph{Phys. Rev.}
  {\bfseries D66} (2002) 023004}
  [\href{https://arxiv.org/abs/hep-ph/0203221}{{\ttfamily hep-ph/0203221}}].

\bibitem{Abazajian:2016yjj}
{\scshape CMB-S4} collaboration, \emph{{CMB-S4 Science Book, First Edition}},
  \href{https://arxiv.org/abs/1610.02743}{{\ttfamily 1610.02743}}.

\bibitem{Abazajian:2019eic}
K.~Abazajian et~al., \emph{{CMB-S4 Science Case, Reference Design, and Project
  Plan}},  \href{https://arxiv.org/abs/1907.04473}{{\ttfamily 1907.04473}}.

\bibitem{Brust:2013xpv}
C.~Brust, D.E.~Kaplan and M.T.~Walters, \emph{{New Light Species and the CMB}},
  \href{https://doi.org/10.1007/JHEP12(2013)058}{\emph{JHEP} {\bfseries 12}
  (2013) 058} [\href{https://arxiv.org/abs/1303.5379}{{\ttfamily 1303.5379}}].

\bibitem{Salvio:2013iaa}
A.~Salvio, A.~Strumia and W.~Xue, \emph{{Thermal axion production}},
  \href{https://doi.org/10.1088/1475-7516/2014/01/011}{\emph{JCAP} {\bfseries
  1401} (2014) 011} [\href{https://arxiv.org/abs/1310.6982}{{\ttfamily
  1310.6982}}].

\bibitem{Baumann:2016wac}
D.~Baumann, D.~Green and B.~Wallisch, \emph{{New Target for Cosmic Axion
  Searches}}, \href{https://doi.org/10.1103/PhysRevLett.117.171301}{\emph{Phys.
  Rev. Lett.} {\bfseries 117} (2016) 171301}
  [\href{https://arxiv.org/abs/1604.08614}{{\ttfamily 1604.08614}}].

\bibitem{Ferreira:2018vjj}
R.Z.~Ferreira and A.~Notari, \emph{{Observable Windows for the QCD Axion
  Through the Number of Relativistic Species}},
  \href{https://doi.org/10.1103/PhysRevLett.120.191301}{\emph{Phys. Rev. Lett.}
  {\bfseries 120} (2018) 191301}
  [\href{https://arxiv.org/abs/1801.06090}{{\ttfamily 1801.06090}}].

\bibitem{DEramo:2018vss}
F.~D'Eramo, R.Z.~Ferreira, A.~Notari and J.L.~Bernal, \emph{{Hot Axions and the
  $H_0$ tension}},
  \href{https://doi.org/10.1088/1475-7516/2018/11/014}{\emph{JCAP} {\bfseries
  1811} (2018) 014} [\href{https://arxiv.org/abs/1808.07430}{{\ttfamily
  1808.07430}}].

\bibitem{Aprile:2020tmw}
{\scshape XENON} collaboration, \emph{{Excess electronic recoil events in
  XENON1T}}, \href{https://doi.org/10.1103/PhysRevD.102.072004}{\emph{Phys.
  Rev. D} {\bfseries 102} (2020) 072004}
  [\href{https://arxiv.org/abs/2006.09721}{{\ttfamily 2006.09721}}].

\bibitem{Arias-Aragon:2020qtn}
F.~Arias-Aragon, F.~D'Eramo, R.Z.~Ferreira, L.~Merlo and A.~Notari,
  \emph{{Cosmic Imprints of XENON1T Axions}},
  \href{https://doi.org/10.1088/1475-7516/2020/11/025}{\emph{JCAP} {\bfseries
  11} (2020) 025} [\href{https://arxiv.org/abs/2007.06579}{{\ttfamily
  2007.06579}}].

\bibitem{Bernal:2016gxb}
J.L.~Bernal, L.~Verde and A.G.~Riess, \emph{{The trouble with $H_0$}},
  \href{https://doi.org/10.1088/1475-7516/2016/10/019}{\emph{JCAP} {\bfseries
  1610} (2016) 019} [\href{https://arxiv.org/abs/1607.05617}{{\ttfamily
  1607.05617}}].

\bibitem{Georgi:1986df}
H.~Georgi, D.B.~Kaplan and L.~Randall, \emph{{Manifesting the Invisible Axion
  at Low-Energies}},
  \href{https://doi.org/10.1016/0370-2693(86)90688-X}{\emph{Phys. Lett.}
  {\bfseries B169} (1986) 73}.

\bibitem{Kaplan:1985dv}
D.B.~Kaplan, \emph{{Opening the Axion Window}},
  \href{https://doi.org/10.1016/0550-3213(85)90319-0}{\emph{Nucl. Phys.}
  {\bfseries B260} (1985) 215}.

\bibitem{Srednicki:1985xd}
M.~Srednicki, \emph{{Axion Couplings to Matter. 1. CP Conserving Parts}},
  \href{https://doi.org/10.1016/0550-3213(85)90054-9}{\emph{Nucl. Phys.}
  {\bfseries B260} (1985) 689}.

\bibitem{Bardeen:1986yb}
W.A.~Bardeen, R.D.~Peccei and T.~Yanagida, \emph{{Constraints on Variant Axion
  Models}}, \href{https://doi.org/10.1016/0550-3213(87)90003-4}{\emph{Nucl.
  Phys.} {\bfseries B279} (1987) 401}.

\bibitem{Fields:2019pfx}
B.D.~Fields, K.A.~Olive, T.-H.~Yeh and C.~Young, \emph{{Big-Bang
  Nucleosynthesis After Planck}},
  \href{https://doi.org/10.1088/1475-7516/2020/03/010}{\emph{JCAP} {\bfseries
  03} (2020) 010} [\href{https://arxiv.org/abs/1912.01132}{{\ttfamily
  1912.01132}}].

\bibitem{Aghanim:2018eyx}
{\scshape Planck} collaboration, \emph{{Planck 2018 results. VI. Cosmological
  parameters}},  \href{https://arxiv.org/abs/1807.06209}{{\ttfamily
  1807.06209}}.

\bibitem{Arik:2008mq}
{\scshape CAST} collaboration, \emph{{Probing eV-scale axions with CAST}},
  \href{https://doi.org/10.1088/1475-7516/2009/02/008}{\emph{JCAP} {\bfseries
  0902} (2009) 008} [\href{https://arxiv.org/abs/0810.4482}{{\ttfamily
  0810.4482}}].

\bibitem{Irastorza:2011gs}
I.G.~Irastorza et~al., \emph{{Towards a New Generation Axion Helioscope}},
  \href{https://doi.org/10.1088/1475-7516/2011/06/013}{\emph{JCAP} {\bfseries
  1106} (2011) 013} [\href{https://arxiv.org/abs/1103.5334}{{\ttfamily
  1103.5334}}].

\bibitem{Armengaud:2014gea}
E.~Armengaud et~al., \emph{{Conceptual Design of the International Axion
  Observatory (IAXO)}},
  \href{https://doi.org/10.1088/1748-0221/9/05/T05002}{\emph{JINST} {\bfseries
  9} (2014) T05002} [\href{https://arxiv.org/abs/1401.3233}{{\ttfamily
  1401.3233}}].

\bibitem{MartinCamalich:2020dfe}
J.~Martin~Camalich, M.~Pospelov, P.N.H.~Vuong, R.~Ziegler and J.~Zupan,
  \emph{{Quark Flavor Phenomenology of the QCD Axion}},
  \href{https://doi.org/10.1103/PhysRevD.102.015023}{\emph{Phys. Rev. D}
  {\bfseries 102} (2020) 015023}
  [\href{https://arxiv.org/abs/2002.04623}{{\ttfamily 2002.04623}}].

\bibitem{Gorghetto:2018myk}
M.~Gorghetto, E.~Hardy and G.~Villadoro, \emph{{Axions from Strings: the
  Attractive Solution}},
  \href{https://doi.org/10.1007/JHEP07(2018)151}{\emph{JHEP} {\bfseries 07}
  (2018) 151} [\href{https://arxiv.org/abs/1806.04677}{{\ttfamily
  1806.04677}}].

\bibitem{Gorghetto:2020qws}
M.~Gorghetto, E.~Hardy and G.~Villadoro, \emph{{More Axions from Strings}},
  \href{https://arxiv.org/abs/2007.04990}{{\ttfamily 2007.04990}}.

\bibitem{Arias-Aragon:2017eww}
F.~Arias-Aragon and L.~Merlo, \emph{{The Minimal Flavour Violating Axion}},
  \href{https://doi.org/10.1007/JHEP10(2017)168}{\emph{JHEP} {\bfseries 10}
  (2017) 168} [\href{https://arxiv.org/abs/1709.07039}{{\ttfamily
  1709.07039}}].

\bibitem{DiLuzio:2017pfr}
L.~Di~Luzio, F.~Mescia and E.~Nardi, \emph{{The Window for Preferred Axion
  Models}},  \href{https://arxiv.org/abs/1705.05370}{{\ttfamily 1705.05370}}.

\bibitem{Chivukula:1987py}
R.~Chivukula and H.~Georgi, \emph{{Composite Technicolor Standard Model}},
  \href{https://doi.org/10.1016/0370-2693(87)90713-1}{\emph{Phys. Lett. B}
  {\bfseries 188} (1987) 99}.

\bibitem{DAmbrosio:2002vsn}
G.~D'Ambrosio, G.~Giudice, G.~Isidori and A.~Strumia, \emph{{Minimal flavor
  violation: An Effective field theory approach}},
  \href{https://doi.org/10.1016/S0550-3213(02)00836-2}{\emph{Nucl. Phys. B}
  {\bfseries 645} (2002) 155}
  [\href{https://arxiv.org/abs/hep-ph/0207036}{{\ttfamily hep-ph/0207036}}].

\bibitem{Vilenkin:1982ks}
A.~Vilenkin and A.~Everett, \emph{{Cosmic Strings and Domain Walls in Models
  with Goldstone and PseudoGoldstone Bosons}},
  \href{https://doi.org/10.1103/PhysRevLett.48.1867}{\emph{Phys. Rev. Lett.}
  {\bfseries 48} (1982) 1867}.

\bibitem{Vachaspati:1984yi}
T.~Vachaspati, A.E.~Everett and A.~Vilenkin, \emph{{Radiation From Vacuum
  Strings and Domain Walls}},
  \href{https://doi.org/10.1103/PhysRevD.30.2046}{\emph{Phys. Rev. D}
  {\bfseries 30} (1984) 2046}.

\bibitem{Chang:1998tb}
S.~Chang, C.~Hagmann and P.~Sikivie, \emph{{Studies of the motion and decay of
  axion walls bounded by strings}},
  \href{https://doi.org/10.1103/PhysRevD.59.023505}{\emph{Phys. Rev. D}
  {\bfseries 59} (1999) 023505}
  [\href{https://arxiv.org/abs/hep-ph/9807374}{{\ttfamily hep-ph/9807374}}].

\bibitem{Hagmann:2000ja}
C.~Hagmann, S.~Chang and P.~Sikivie, \emph{{Axion radiation from strings}},
  \href{https://doi.org/10.1103/PhysRevD.63.125018}{\emph{Phys. Rev. D}
  {\bfseries 63} (2001) 125018}
  [\href{https://arxiv.org/abs/hep-ph/0012361}{{\ttfamily hep-ph/0012361}}].

\bibitem{Vilenkin:1981zs}
A.~Vilenkin, \emph{{Gravitational Field of Vacuum Domain Walls and Strings}},
  \href{https://doi.org/10.1103/PhysRevD.23.852}{\emph{Phys. Rev. D} {\bfseries
  23} (1981) 852}.

\bibitem{Sikivie:1982qv}
P.~Sikivie, \emph{{Of Axions, Domain Walls and the Early Universe}},
  \href{https://doi.org/10.1103/PhysRevLett.48.1156}{\emph{Phys. Rev. Lett.}
  {\bfseries 48} (1982) 1156}.

\bibitem{Ipser:1983db}
J.~Ipser and P.~Sikivie, \emph{{The Gravitationally Repulsive Domain Wall}},
  \href{https://doi.org/10.1103/PhysRevD.30.712}{\emph{Phys. Rev. D} {\bfseries
  30} (1984) 712}.

\bibitem{Vilenkin:1984hy}
A.~Vilenkin, \emph{{Gravitational Field of Vacuum Domain Walls}},
  \href{https://doi.org/10.1016/0370-2693(83)90554-3}{\emph{Phys. Lett. B}
  {\bfseries 133} (1983) 177}.

\bibitem{Calibbi:2016hwq}
L.~Calibbi, F.~Goertz, D.~Redigolo, R.~Ziegler and J.~Zupan, \emph{{Minimal
  axion model from flavor}},
  \href{https://doi.org/10.1103/PhysRevD.95.095009}{\emph{Phys. Rev. D}
  {\bfseries 95} (2017) 095009}
  [\href{https://arxiv.org/abs/1612.08040}{{\ttfamily 1612.08040}}].

\bibitem{Ema:2016ops}
Y.~Ema, K.~Hamaguchi, T.~Moroi and K.~Nakayama, \emph{{Flaxion: a minimal
  extension to solve puzzles in the standard model}},
  \href{https://doi.org/10.1007/JHEP01(2017)096}{\emph{JHEP} {\bfseries 01}
  (2017) 096} [\href{https://arxiv.org/abs/1612.05492}{{\ttfamily
  1612.05492}}].

\bibitem{Millea:2015qra}
M.~Millea, L.~Knox and B.~Fields, \emph{{New Bounds for Axions and Axion-Like
  Particles with Kev-Gev Masses}},
  \href{https://doi.org/10.1103/PhysRevD.92.023010}{\emph{Phys. Rev.}
  {\bfseries D92} (2015) 023010}
  [\href{https://arxiv.org/abs/1501.04097}{{\ttfamily 1501.04097}}].

\bibitem{Feng:1997tn}
J.L.~Feng, T.~Moroi, H.~Murayama and E.~Schnapka, \emph{{Third Generation
  Familons, B Factories, and Neutrino Cosmology}},
  \href{https://doi.org/10.1103/PhysRevD.57.5875}{\emph{Phys. Rev.} {\bfseries
  D57} (1998) 5875} [\href{https://arxiv.org/abs/hep-ph/9709411}{{\ttfamily
  hep-ph/9709411}}].

\bibitem{Keller:2012yr}
J.~Keller and A.~Sedrakian, \emph{{Axions from Cooling Compact Stars}},
  \href{https://doi.org/10.1016/j.nuclphysa.2012.11.004}{\emph{Nucl. Phys.}
  {\bfseries A897} (2013) 62}
  [\href{https://arxiv.org/abs/1205.6940}{{\ttfamily 1205.6940}}].

\bibitem{Sedrakian:2015krq}
A.~Sedrakian, \emph{{Axion Cooling of Neutron Stars}},
  \href{https://doi.org/10.1103/PhysRevD.93.065044}{\emph{Phys. Rev.}
  {\bfseries D93} (2016) 065044}
  [\href{https://arxiv.org/abs/1512.07828}{{\ttfamily 1512.07828}}].

\bibitem{Fischer:2016cyd}
T.~Fischer, S.~Chakraborty, M.~Giannotti, A.~Mirizzi, A.~Payez and A.~Ringwald,
  \emph{{Probing Axions with the Neutrino Signal from the Next Galactic
  Supernova}}, \href{https://doi.org/10.1103/PhysRevD.94.085012}{\emph{Phys.
  Rev.} {\bfseries D94} (2016) 085012}
  [\href{https://arxiv.org/abs/1605.08780}{{\ttfamily 1605.08780}}].

\bibitem{Giannotti:2017hny}
M.~Giannotti, I.G.~Irastorza, J.~Redondo, A.~Ringwald and K.~Saikawa,
  \emph{{Stellar Recipes for Axion Hunters}},
  \href{https://doi.org/10.1088/1475-7516/2017/10/010}{\emph{JCAP} {\bfseries
  1710} (2017) 010} [\href{https://arxiv.org/abs/1708.02111}{{\ttfamily
  1708.02111}}].

\bibitem{Aubert:2004ws}
{\scshape BaBar} collaboration, \emph{{A search for the decay $B^+ \to K^+ \nu
  \bar{\nu}$}},
  \href{https://doi.org/10.1103/PhysRevLett.94.101801}{\emph{Phys. Rev. Lett.}
  {\bfseries 94} (2005) 101801}
  [\href{https://arxiv.org/abs/hep-ex/0411061}{{\ttfamily hep-ex/0411061}}].

\bibitem{UTfit}
\emph{{Updated fits of the UTfit collaboration at
  \url{http://www.utfit.org/UTfit/}}}, .

\bibitem{Lees:2013kla}
{\scshape BaBar} collaboration, \emph{{Search for $B \to K^{(*)} \nu \overline
  \nu$ and invisible quarkonium decays}},
  \href{https://doi.org/10.1103/PhysRevD.87.112005}{\emph{Phys. Rev. D}
  {\bfseries 87} (2013) 112005}
  [\href{https://arxiv.org/abs/1303.7465}{{\ttfamily 1303.7465}}].

\bibitem{Adler:2008zza}
{\scshape E949, E787} collaboration, \emph{{Measurement of the K+
  --\ensuremath{>} pi+ nu nu branching ratio}},
  \href{https://doi.org/10.1103/PhysRevD.77.052003}{\emph{Phys. Rev. D}
  {\bfseries 77} (2008) 052003}
  [\href{https://arxiv.org/abs/0709.1000}{{\ttfamily 0709.1000}}].

\end{thebibliography}\endgroup

\end{document}